\documentclass[lettersize,journal]{IEEEtran}
\usepackage{amsmath,amsfonts}
\usepackage{algorithmic}
\usepackage{algorithm}
\usepackage{array}
\usepackage[caption=false,font=footnotesize]{subfig} 
\usepackage{textcomp}
\usepackage{stfloats}
\usepackage{url}
\usepackage{verbatim}
\usepackage{graphicx}
\usepackage{cite}
\usepackage{hyperref}

\usepackage{gensymb}

\usepackage{multirow}
\usepackage{xcolor}
\usepackage{cuted}
\usepackage{amsthm}
\usepackage{mathtools}

\newtheorem{proposition}{Proposition}
\newtheorem{assumption}{Assumption}

\usepackage{amsmath,amsfonts,bm}

\def\eqref#1{equation~\ref{#1}}

\def\1{\bm{1}}

\def\vd{{\bm{d}}}

\def\vg{{\bm{g}}}
\def\vh{{\bm{h}}}

\def\vu{{\bm{u}}}
\def\vv{{\bm{v}}}

\def\vy{{\bm{y}}}
\def\vz{{\bm{z}}}

\def\mE{{\bm{E}}}

\def\mI{{\bm{I}}}

\def\mU{{\bm{U}}}

\DeclareMathAlphabet{\mathsfit}{\encodingdefault}{\sfdefault}{m}{sl}
\SetMathAlphabet{\mathsfit}{bold}{\encodingdefault}{\sfdefault}{bx}{n}

\def\sC{{\mathbb{C}}}

\newcommand{\sSigma}{\widehat{\mathbf{\Sigma}}}

\DeclareMathOperator*{\argmax}{arg\,max}

\usepackage{xr}
\makeatletter

\newcommand*{\addFileDependency}[1]{
\typeout{(#1)}
\@addtofilelist{#1}
\IfFileExists{#1}{}{\typeout{No file #1.}}
}\makeatother

\begin{document}

\title{Domain Adaptation for DoA Estimation in Multipath Channels with Interferences
}

\author{
Amitay Bar, Joseph S. Picard, Israel Cohen, and Ronen Talmon
\thanks{The authors are with the Viterbi Faculty of Electrical and Computer Engineering, Technion—-Israel Institute of Technology, Haifa 32000, Israel (e-mail: amitayb@campus.technion.ac.il; picard.joseph.post@gmail.com; icohen@ee.technion.ac.il; ronen@ef.technion.ac.il). This work was supported by the European Union’s Horizon 2020 research and innovation program under grant agreement No. 802735-ERC-DIFFOP.}
}



\maketitle

\begin{abstract}
We consider the problem of estimating the direction-of-arrival (DoA) of a desired source located in a known region of interest in the presence of interfering sources and multipath.
We propose an approach that precedes the DoA estimation and relies on generating a set of reference steering vectors. The steering vectors' generative model is a free space model, which is beneficial for many DoA estimation algorithms. The set of reference steering vectors is then used to compute a function that maps the received signals from the adverse environment to a reference domain free from interfering sources and multipath. 
We show theoretically and empirically that the proposed map, which is analogous to domain adaption, improves DoA estimation by mitigating interference and multipath effects. Specifically, we demonstrate a substantial improvement in accuracy when the proposed approach is applied before three commonly used beamformers: the delay-and-sum (DS), the minimum variance distortionless response (MVDR), and the Multiple Signal Classification (MUSIC).

\end{abstract}


\section{Introduction}
\IEEEPARstart{E}{stimation} of the Direction of Arrival (DoA) of a desired source is a longstanding problem. Despite being extensively investigated in the last decades, it is still considered an active field that attracts attention and interest. DoA estimation has numerous applications in a broad range of fields, such as automated camera steering, teleconferencing systems \cite{huang2000passive}, speaker separation \cite{mandel2009model}, radar \cite{li2007mimo}, smart sensor arrays for 5G networks \cite{abdelbari2020novel}, and autonomous vehicles \cite{wan2020deep}, to name but a few. 

Currently, two of the major challenges in DoA estimation are the presence of interfering sources and multipath channels.
Interfering sources could mask the desired source, while multipath channels between the desired source and the phased array could cause deviations in the DoA estimation of the desired source. 
A central approach to overcoming these challenges is learning the environment before the localization. 
For example, such an approach was first introduced in acoustic source localization in \cite{malioutov2005sparse}, where a predefined grid of potential source positions was considered.
In the context of communication signals, a similar approach based on fingerprinting has been proposed in \cite{wax2001radio,ahonen2003database,kupershtein2012single}.
More recently, in \cite{xiao2015learning}, such an approach was used to mitigate the multipath that stems from a reverberant environment by a learning-based method that formulates DoA estimation as a classification problem. 
A localization approach based on manifold learning has also been proposed in \cite{talmon2011supervised,laufer2017semi,laufer2016semi}.
However, all these methods inherently depend on the environment and require labeled samples, i.e., signal measurements associated with a known DoA. 
Another recent approach is based on Riemannian geometry %
\cite{10283470}. There, interfering sources located in a reverberant environment are considered, and an approach for beamformer design and interference rejection based on Riemannian geometry is proposed. 
However, the interfering sources are assumed to be only partially active.
{Methods for superresolution have also been proposed \cite{han2022two,zhu2019multi}. However, for the purpose of interference rejection they require the identification of the different sources, which in turn, requires prior knowledge that is not always at hand.}
In recent years, with the development of Deep Learning (DL), a multitude of DL-based methods have been introduced for robust DoA estimation \cite{chakrabarty2019multi,huang2018deep,chen2020deep,ozanich2020feedforward,vera2021acoustic,grumiaux2022survey}. 
Various variants, such as time-frequency masks incorporated into DL methods, have also been proposed to mitigate noise and interference \cite{mack2020signal,mack2022signal,wang2018robust}.
However, DL-based methods typically require an extensive data set for training the model, which is not always at hand. Furthermore, DL methods usually exhibit higher computational complexity than classical signal processing methods.

In this paper, we consider the problem of DoA estimation of a desired source located in a known region of interest in the presence of interfering sources and multipaths. 
Leveraging the prior knowledge of the known region of interest and the available location of the phased array, {we are able to mitigate the adverse condition without requiring any labeled samples (prerecorded signals). Inspired by domain adaptation in machine learning and parallel transport in Riemannian geometry, }we generate a set of simulated steering vectors using a free space model. 
This set is used to build a map applied to the received signals, mapping them from the adverse environment to a reference environment, free from interference and multipath.
Subsequently, in the reference environment, we employ the DoA estimation algorithm of choice for the mapped signals rather than the received ones. %

Broadly, our approach introduces two new viewpoints to DoA estimation. First, our approach is analogous to domain adaptation (DA), a long-standing problem in machine learning that has been extensively studied \cite{blitzer2007learning,ben2006analysis,farahani2021brief_arXiv,wang2018deep,you2019universal}. We consider complex steering vectors lying in the Euclidean space $\mathbb{C}^M$, where $M$ is the number of elements in the phased array, in which we designate two domains. The reference domain is defined by the simulated steering vectors and represents a free space environment without multipaths and interfering sources. The operational domain is determined by the received (noisy) signals. We devise a map that transforms a received signal from the operational domain to the reference domain, mitigating multipath and interference effects.

From a geometric standpoint, we examine the sample correlation matrices of the received noisy signals and consider them together with the model correlation matrices stemming from the reference steering vectors in the space of $M \times M$ Hermitian positive definite (HPD) matrices. We show that the map we built could be described analogously as parallel transport in the manifold of HPD matrices equipped with the Affine Invariant metric \cite{pennec2006riemannian,yair2019parallel,hiai2009riemannian,bhatia2006riemannian,nielsen2013matrix}.
This map alleviates an inherent shortcoming of prototypical DoA estimation algorithms designed for ideal covariance matrices, e.g., population covariance matrices stemming from a free space with only a single source without interference or multipath. However, these covariance matrices are often not available in practice. 
We, therefore, propose an approach for mapping the real-world covariance matrices from the operational domain to a reference domain that represents ideal covariance matrices. 

{The main contributions of the papers are as follows. First, we introduce a new approach for DoA estimation in adverse conditions focusing on interference rejection. The proposed approach learns the environment without requiring prerecorded signals or prior access to it, making it suitable for various applications. The proposed approach results in a linear map that exhibits low complexity and is mathematically tractable. Additionally, it is a general approach that can be applied to multiple DoA estimation methods.}

{Second, we experimentally demonstrate that the proposed approach improves DoA estimation in environments with interfering sources and multipaths when using baseline beamformers. In the experiments, we considered both acoustic signals and radio frequency (RF) signals, demonstrating the applicability of our approach to a wide range of applications.
While our approach is general and fits a multitude of downstream DoA estimation methods, our focus is on the following three standard beamformers: the Delay and Sum (DS) beamformer \cite{krim1996two,stoica2005spectral}, the
Minimum Variance Distortionless Response (MVDR) beamformer \cite{capon1969high}, and the Multiple Signal Classification (MUSIC) beamformer \cite{schmidt1986multiple,yan2013low,zhang2010direction}. 
These three beamformers are considered because they are widely used and demonstrate different approaches for DoA estimation: a statistical approach (DS), minimizing an objective function (MVDR), and a subspace approach (MUSIC). 
Importantly, our approach is not limited to these beamformers; we also apply it to the method presented in \cite{amar2015linearly} for beam pattern design with a null section, and show that it leads to improved results. 
Moreover, we examine the SRP of the proposed approach combined with the DS beamformer and show that it implicitly includes nulls in the direction of the interfering sources, implying that the energy of the sample correlation matrix in the directions of the interfering sources is reduced.}

{Third, we provide a theoretical analysis of the proposed approach.
We show that the map we built matches the second-order statistics of the mapped signals with that of the steering vectors in the reference environment, representing the model typically considered by DoA estimation methods.
Additionally, we show that for equipower desired and interfering sources, and under some conditions, the SRP of the DS beamformer is higher in the direction of the desired source than in the direction of the interfering source. }

The remainder of this paper is structured as follows. In Section \ref{sec: Problem Formulation}, we formulate the problem. Section \ref{sec: Background on } presents a brief background on the considered beamformers. In Section \ref{Sec: Proposed Approach}, we describe and analyze the proposed approach. In addition, we relate the proposed approach to concepts from Riemannian geometry. Section \ref{Sec: simulation results} presents the simulation results. We conclude the work in Section \ref{sec: Conclusion}.

\section{Problem Formulation} 
\label{sec: Problem Formulation}

We consider the problem of estimating the DoA of a desired source using a phased array. Suppose the desired source transmits from various positions within a known region of interest at different times. 
In addition to the desired source, suppose moving interfering sources transmit from multiple locations within an unknown interference region. We assume that the region of interest and the region of interference do not overlap. We also consider the possibility of additional, fixed-location interfering sources within or outside the region of interest.

The source signals are received at an array of $M$ elements (acoustic microphones or RF antennas) positioned in known locations. 
The received signal at the $m$th element is given by
\begin{equation}
    z_m(n) =
    s(n)*h_m(n) +
    \sum_{i=1}^{N_I}q_i(n)*g_{im}(n) +
    v_m(n),
\end{equation}
where $s(n)$ is the signal of the desired source, $q_i(n)$ is the signal of the $i$th interfering source, $N_I$ is the number of currently active interfering sources, $h_m(n)$ is the impulse response of the channel between the source and the $m$th element, $g_{im}(n)$ is the impulse response of the channel between the $i$th interfering source and the $m$th element, and $v_m(n)$ is the noise at the $m$ element.
{The noise is assumed to be spatially white, and the interfering sources are assumed to be uncorrelated with the desired sources. }

Our goal is to estimate the DoA of the desired source. The main challenges in the considered setting are interfering sources, which could be stronger than the desired source, and the indirect paths in the channel between the desired source and the phased array.

{We focus on narrowband signals and analyze the received signal in the frequency domain.
In section \ref{Sec: simulation results}, we consider acoustic sources and discuss the extension to wideband signals.
Let $z_m(l)$ denote the received signal at the frequency domain at the $l$th time, which is given by
\begin{equation}
\label{eq: received signal model z_m(l,k)}
    z_m(l) =
    s(l)h_m(l) +
    \sum_{i=1}^{N_I}q_i(l)g_{im}(l) +
    v_m(l),
\end{equation}
where the notation for $s(l),h_m(l),q_i(l),g_{im}(l)$ and $v_m(l)$ follows similarly.
We analyze a finite interval of the received signal and denote the number of available samples by $L$.
}

We stack the received signal from all the elements in the array in a vector form to obtain
\begin{equation}
\label{eq: received signal z as vector}
    \vz(l) = 
    [z_1(l)\;\;z_2(l)\;\cdots \; z_M(l)]^\top
    \in\sC^{M\times 1}.
\end{equation}
Similarly, the noise term in vector form is given by
\begin{equation}
    \vv(l) = 
    [v_1(l)\;\;v_2(l)\;\cdots \; v_M(l)]^\top
    \in\sC^{M\times 1}.
\end{equation}
We assume that the impulse response of the channels between the sources and the phased array does not change over time. Thus, we omit their index $l$. Consequently, their vector form is given by 
\begin{equation}
    \vh = 
    [h_1\;\;h_2\;\cdots \; h_M]^\top
    \in\sC^{M\times 1},
\end{equation}
and 
\begin{equation}
    \vg_i = 
    [g_{i1}\;\;g_{i2}\;\cdots \; g_{iM}]^\top
    \in\sC^{M\times 1}.
\end{equation}
The received signal can be explicitly expressed as 
\begin{equation}
    \vz(l) = 
    s(l)\vh + 
    \sum_{i=1}^{N_I}
    q_i(l)\vg_i + 
    \vv(l).
\end{equation}
{We emphasize that the statistical properties of the interfering sources are arbitrary. Importantly, the proposed approach does not require prior knowledge or specific statistical properties.}

We conclude this section with two remarks.
First, the formulation supports various applications, including acoustic and RF. 
Second, instead of a single desired source transmitting from various positions, multiple desired sources that are not simultaneously active could exist. %

\section{Background on Beamformers}
\label{sec: Background on }

Let $\vd(\theta)$ denote the steering vector of the array to direction $\theta$, which is given by
\begin{equation}
    \vd(\theta) = [1,e^{j\phi_2(\theta)},...,e^{j\phi_M(\theta)}]^\top \;\;\; \in\sC^{M\times 1},
\end{equation}
where $\phi_m(\theta)$ is the phase of the received signal at the $m$th element with respect to a reference element. For example, the phase at the $m$th element for a uniform linear array is given by \cite{krim1996two} $\phi_m(\theta) = 2\pi m\frac{\delta}{\lambda}\sin\theta$, where $\delta$ is the distance between the elements, $\lambda$ is the wavelength of the received signal, and the typical indexing is considered.

In this paper, we focus on three widely used beamformers: the DS beamformer \cite{krim1996two}, MVDR \cite{capon1969high}, and MUSIC \cite{schmidt1986multiple}. They are all based on the sample correlation matrix of the received signal, denoted by $\sSigma\in\sC^{M\times M}$ and given by
\begin{equation}
    \sSigma = 
    \frac{1}{L}\sum_{l=1}^L \vz(l)\vz^H(l),
\end{equation}
where $(\cdot)^H$ is the conjugate transpose operator.
The spectrum of the DS beamformer is given by \cite{widrow1967adaptive,compton1988adaptive}
\begin{equation}
    P_{\text{DS}}(\sSigma,\theta) = 
    \vd^H(\theta)\sSigma\vd(\theta),
\end{equation}
and the spectrum of the MVDR beamformer is given by \cite{capon1969high}
\begin{equation}
    P_{\text{MVDR}}(\sSigma,\theta) = 
    \frac{1}{\vd^H(\theta)\sSigma^{-1}\vd(\theta)}.
\end{equation}
For MUSIC \cite{schmidt1986multiple}, the spectral decomposition of $\sSigma$ is performed, and
the eigenvectors of $\sSigma$ associated with the noise are extracted. They form the columns of the matrix spanning the noise subspace, denoted by $\mU_N$. 
The spectrum associated with MUSIC is given by 
\begin{equation}
    P_{\text{MUSIC}}(\sSigma,\theta) = 
    \frac{1}{\vd^H(\theta)\mU_N\mU_N^H\vd(\theta)}.
\end{equation}
For all three beamformers, given a sample correlation matrix $\sSigma$, the DoA estimation of a single source is realized by 
\begin{equation}
\label{eq: theta hat = argmax P background}
    \begin{split}
        \hat{\theta} = 
        \argmax_\theta P(\sSigma,\theta),
    \end{split}
\end{equation}
where $P(\sSigma,\theta)$ is the spectrum of the respective beamformer.

\section{Proposed Approach}
\label{Sec: Proposed Approach}
We begin with a brief overview of the proposed approach and a detailed algorithm description.

\subsection{Overview}

The main idea underlying our approach is that knowledge of the phased array's location, constellation, and information about the region of interest allows us to generate simulated steering vectors. 
{We emphasize that these steering vectors are generated analytically, allowing a specific design to improve the DoA estimation.}
We postulate that these precomputed steering vectors facilitate the improvement of the DoA estimation after deploying the array in the operational environment consisting of multipath and undesired interfering sources.

More concretely, before deploying the phased array, 
generating steering vectors with appropriate attenuation from various positions within the known region of interest. 
We remark that these steering vectors could be viewed as transfer functions (TFs) of a reference environment considering only the direct path. We refer to them as simulated steering vectors, even though they also consist of an attenuation term.

After deploying the array, the received signals are collected in a time segment called the \emph{adaptation phase}. 
These signals are viewed as measurements characterizing the operational environment. 
Combined with the simulated steering vectors, these measurements are used to compute a map of received signals in the operational environment, with interfering sources and channels with indirect paths, to the reference environment, consisting of only the direct path and without interferences. This map is then applied to the subsequently received signals.

Although commonly used beamformers might not provide accurate results due to interfering sources and indirect paths, our approach applies the beamformers to the mapped, adapted signals rather than the received ones.
Through this adaptation, the signals become closer to the reference environment, attenuating the interfering sources and the effects of the indirect paths.
A detailed description of the algorithm is presented in the next section.

We note that the phased array is deployed in the operational environment during the adaptation phase. To avoid interfering with the system's operation, the DoA can still be estimated based on the received signals using standard beamformers.
Note that throughout the paper, we use the terms `domain' and `environment' interchangeably according to the context.

 \subsection{Algorithm}

{A key component of the proposed approach is steering vectors. }
The algorithm's first step is to generate the simulated steering vectors from the reference environment. This is done only once before the deployment of the phased array.
{We consider a generative model describing a reference environment that is more adequate for the subsequent DoA estimation algorithm and is typically simpler than the operational environment. The steering vector generated using the model is denoted by
\begin{equation}
\label{eq: TF as vector}
    \tilde{\vd}(\theta,r_1,\ldots,r_M) = [\tilde{d}_1(\theta,r_1) \cdots \tilde{d}_M(\theta,r_M)]^\top \in \sC^{M\times 1},
\end{equation}
where $r_m$ is the distance between the source and the $m$th element, and $\tilde{d}_m$ is a complex value scalar describing the channel between the source and the $m$th element.
In this paper, we consider standard beamformer, so we use the free space model \cite{krim1996two} as follows 
\begin{equation}
\label{eq: TF for FS}
    \tilde{d}_m = \frac{1}{r_m}e^{-j2\pi \frac{r_m}{\lambda}}.
\end{equation}
We note that our approach is general, and other models could also be considered. }
The model in (\ref{eq: TF for FS}) is used to generate $N_{\text{S}}$ different steering vectors from $N_{\text{S}}$ different positions, arbitrarily chosen, in the region of interest. We denote by $\tilde{\vd}_j$ the steering vector associated with the $j$th position.

Next, we compute the correlation matrix of each steering vector as follows. 
\begin{equation}
\label{eq: Sigma^A_j}
    \begin{split}     
         \mathbf{\Sigma}_j^{\text{S}} &=         
        \tilde{\vd}_j\tilde{\vd}_j^H + 
        \epsilon\mI \;\;\;\;
        j=1,\ldots,N_{\text{S}},
    \end{split}
\end{equation}
where $\epsilon$ is a small positive constant, and the term $\epsilon \mI$ is for numerical stability (which could also be viewed as a noise term considering high SNR). We note that the empirical results are insensitive to the choice of $\epsilon$, even for orders of magnitude. We set $\epsilon = 10^{-7}$ in our experiments.

We follow by computing the mean correlation matrix of the $N_{\text{S}}$ correlation matrices, $\{ \mathbf{\Sigma}_j^{\text{S}}\}_{j=1}^{N_{\text{S}}}$, which is given by
\begin{equation}
\label{eq: Sigma_A DoA domain}
    \begin{split}    
         \mathbf{\Sigma}_{\text{S}} =      
    \frac{1}{N_{\text{S}}}\sum_{j=1}^{N_{\text{S}}}  \mathbf{\Sigma}_j^{\text{S}}
    .
    \end{split}
\end{equation}
{We emphasize that the matrix $\mathbf{\Sigma}_{\text{S}}$ is created using only simulated steering vectors, without any received signals.}

Next, we turn to the adaptation phase performed immediately after the deployment of the system in the operational environment, during which we collect $N_{\text{A}}$ signals from $N_{\text{A}}$ different positions in the region of interest.
{We assume the interfering sources are active during the adaptation phase}.
The received signal from the $i$th position is denoted by $\vz_i(l)\in\sC^{M\times 1}$, and the set of $N_{\text{A}}$ signals from $N_{\text{A}}$ distinct positions in the region of interest is denoted by $\{\vz_i(l)\}_{i=1}^{N_{\text{A}}}$. 
{We recall that we do not need to know the positions of these transmissions, and in practice, they could be obtained during the normal activity of the desired source.
}

The received signals at the adaptation phase
$\{\vz_i(l)\}_{i=1}^{N_{\text{A}}}$ are used to compute the following sample correlation matrices
\begin{equation}
\label{eq: Sigma^S_i}
    \begin{split}
      \sSigma^{\text{A}}_i &=
        \frac{1}{L}\sum_{l=1}^L \vz_i(l)\vz_i^H (l) \;\; i=1,\ldots,N_{\text{A}}. \\             
    \end{split}
\end{equation}
The mean correlation  matrix of the set $\{ \sSigma^{\text{A}}_i\}_{i=1}^{N_{\text{A}}}$ is computed, and is given by
\begin{equation}
\label{eq: Sigma_S operational domain}
    \begin{split}     
              \mathbf{\Sigma}_{\text{A}} &=      
    \frac{1}{N_{\text{A}}}\sum_{i=1}^{N_{\text{A}}}  \sSigma^{\text{A}}_i.
    \end{split}
\end{equation}

Finally, based on the two mean correlation matrices $ \mathbf{\Sigma}_{\text{S}}$ and $ \mathbf{\Sigma}_{\text{A}}$, we propose the following linear map of the received signal to the reference environment
\begin{equation}
    \vy(l) = \mE\vz(l) \;\; \forall l=1,\ldots ,L,
\end{equation}
where
\begin{equation}
\label{eq: E_DA coral}
    \mE =    
     \mathbf{\Sigma}_{\text{S}}
    ^{\frac{1}{2}}         
    \mathbf{\Sigma}_{\text{A}}
    ^{-\frac{1}{2}} .
\end{equation}
The resulting adapted signal $\vy(l)$ is used to compute the beamformer for the DoA estimation.
The sample correlation matrix of the adapted signal is given by
\begin{equation}
    \begin{split}
    \frac{1}{L}
        \sum_{l=1}^L \vy(l)\vy(l)^H =        \mE\sSigma\mE^H.
    \end{split}
\end{equation}
Since we focus on beamformers, which are computed using the correlation matrices,
the spectrum of the proposed beamformer is computed using $\mE\sSigma\mE^H$ as the correlation matrix, and, instead of (\ref{eq: theta hat = argmax P background}), the DoA of the desired source is estimated by
\begin{equation}
\label{eq: theta hat as argmax}
    \begin{split}
        \hat{\theta} = 
        \argmax_\theta P(\mE\sSigma\mE^H,\theta).
    \end{split}
\end{equation}
where the spectrum $P$ is set to the spectrum of the specific beamformer used, e.g., the DS beamformer, the MVDR beamformer, and MUSIC with spectra denoted by $P_{\text{DS}}(\theta)$, $P_{\text{MVDR}}(\theta)$ and $P_{\text{MUSIC}}(\theta)$, respectively.
{The proposed approach results in interference rejection, allowing for the estimation of the DoA of the desired source using the strongest peak in the spectrum. Furthermore, it avoids the need for source identification.}
We emphasize that other DoA estimation methods that employ the sample correlation matrix can follow the proposed approach.
The algorithm is summarized in Algorithm \ref{Alg: DA for DoA}.
{Following (\ref{eq: theta hat as argmax}), the proposed map is linear, demonstrating low complexity for the proposed approach.}
{While linear maps were considered in the past, the main contribution in the proposed approach is exploiting the knowledge of the region of interest and the position of the phased array, which are typically known, for the construction of a simple map that is both easy to compute and to apply. The map leads to the enhancement of the received signal, which allows more accurate DoA estimates.
We note that the proposed map results in colored noise. However, it also mitigates the effect of the multipath and the interfering sources, which are the main challenges in the considered setting.}

\begin{algorithm}
\caption{The proposed DoA estimation approach}
\label{Alg: DA for DoA}
\textbf{Input:} a set of $N_{\text{A}}$ received signals $\{\vz_i(l)\}_{i=1}^{N_{\text{A}}}$  and a signal from the desired source $\vz(l)$ for $l=0,1,...,L-1$\\
\textbf{Output:} the DoA estimation \\
\begin{algorithmic}[1]
\STATE Generate $N_{\text{S}}$ distinct steering vectors according to (\ref{eq: TF for FS}) and (\ref{eq: TF as vector})
\STATE Compute the set of correlation matrices $\{ \mathbf{\Sigma}_j^{\text{S}}\}_{j=1}^{N_{\text{S}}}$ and $\{ \sSigma_i^{\text{A}}\}_{i=1}^{N_{\text{A}}}$according to (\ref{eq: Sigma^A_j}) and (\ref{eq: Sigma^S_i})
\STATE Compute the mean correlation matrices $ \mathbf{\Sigma}^{\text{S}}$ and $ \mathbf{\Sigma}^{\text{A}}$ according to (\ref{eq: Sigma_A DoA domain}) and (\ref{eq: Sigma_S operational domain})
\STATE Compute the mapping matrix $\mE$ according to (\ref{eq: E_DA coral})
\STATE Estimate the DoA using (\ref{eq: theta hat as argmax})
\end{algorithmic}
\end{algorithm}
The proposed approach can be viewed as DA using CORAL \cite{sun2016return}. 
We view the signals collected after deployment from the adaptation phase $\{\vz_i(l)\}_{i=1}^{N_\text{A}}$ along with the desired signal $\vz(l)$ as belonging to the same domain. This is the operational domain of the operational environment with interfering sources and indirect paths. We view the simulated steering vectors generated analytically before the adaptation phase as stemming from a reference domain, considering a reference environment with only the direct path. 
Adapting the signals by the matrix $\mE$ minimizes the domain shift by aligning the sample correlation matrices \cite{sun2016return}. Specifically, it transports the signal of the desired source from the operational domain, which is typically involved, to the reference domain, which is ``simple''.

We note that for computing the matrix $\mathbf{\Sigma}_{\text{A}}$ in (\ref{eq: Sigma_S operational domain}), it is not required to store the $N_{\text{A}}$ received signals, but the correlation matrix of each signal can be computed and stored instead. 

We conclude with three remarks.
First, the proposed approach could also be used for null steering by designing beamformers with nulls to entire, possibly non-overlapping, sections. See more details in Section \ref{Sec: simulation results}.
Second, the proposed approach is based on the signals received during the adaptation phase. These are unlabeled signals, i.e., the transmitting source's position is unknown. As a result, the proposed approach could be characterized as an unsupervised method.
Third, this paper focuses on the DS, MVDR, and MUSIC beamformers. However, the proposed approach is general and could also be applied to other DoA estimation methods by considering the adapted signals $\vy(l)$ instead of the received signals $\vz(l)$.

\subsection{Relation To Existing Works}
\label{subsec: Relation To Existing Works}
{Applying linear maps to the received signal has been proposed in the past, for example, for array calibration \cite{VIBERG200993}. Seemingly, array calibration is similar to our approach. However, it considers a different setting with a different goal. Array calibration aims at mitigating impairments in the phased array itself, such as misalignment in the elements' position or mutual coupling between the elements \cite{friedlander1991direction,chen2018off}. 
In the considered setting, the main challenge in DoA estimation is the presence of multipath and interfering sources, not the array's impairments. Therefore, we assume a phased array after calibration.
We note that while array calibration methods exist that consider multipath or interferences, they aim to calibrate the array and not reject the interferences for DoA estimation. Additionally, they typically make further assumptions, such as the existence of a calibration source at a known position \cite{sippel2019situ}, knowing the number of interfering sources \cite{he2021simultaneous} or working in near field \cite{sippel2019situ}.}

{Recently, the Bayesian approach has been proposed for the DoA problem through sparse Bayesian learning \cite{hu2016source, dai2021real}. In this framework, the apriori distribution of the weights for the sparse vector is assumed to be known, and the estimated DoAs are according to its conditional distribution.
In contrast, the proposed approach is data-driven. It does not require a priori knowledge of the distribution or assumptions on the statistical model of the different sources.
A key component of the proposed approach involves generating simulated steering vectors that can be viewed as augmented data. Data augmentation for source localization has already been considered in the literature. For example, in \cite{he2021neural}, a DoA estimation method based on neural networks (NNs) is proposed, and DA and augmented data are used to adapt a pre-trained network to the real-world and augmented data. 
Data augmentation has also been proposed in 
\cite{braun2020data} for training an NN for supervised DL-based speech enhancement.
In contrast to \cite{he2021neural,braun2020data}, the proposed approach learns the environment in an \textit{entirely unsupervised} manner, and the adaptation is applied to the received signals directly rather than to an NN.
Furthermore, the present setting also considers interfering sources, as opposed to \cite{he2021neural,braun2020data}.
The proposed approach could be viewed as a data enhancement approach applied before any DoA estimation method. 
}

\subsection{A Riemannian Geometry Perspective}
\label{subsec: A Riemannian Geometry Perspective}

An alternative approach could be based on a geometric standpoint, resulting in a slightly different algorithm implementation.
The sample correlation matrix of the received signal, $\sSigma$, is viewed as a point on the HPD manifold \cite{hiai2009riemannian}, which, equipped with an appropriate metric, forms a Riemannian manifold. 
The point $\sSigma$ is transported along the geodesic path from $ \mathbf{\Sigma}_{\text{A}}$, representing the operational domain, to $ \mathbf{\Sigma}_{\text{S}}$, representing the reference domain, by Parallel Transport (PT) \cite{sra2015conic,yair2019parallel}, which preserves the orientation of $\sSigma$ with respect to the manifold.

The new point after the PT is given by \cite{yair2019parallel}
\begin{equation}
\label{eq: E_pt expression}
    \mathbf{\Sigma}_{PT} =
    \mE_{\text{PT}} \sSigma \mE_{\text{PT}}^H,
\end{equation}
for
\begin{equation}
\label{eq: E_DA pt}
    \mE_{\text{PT}} = \left(  \mathbf{\Sigma}_{\text{S}} \mathbf{\Sigma}_{\text{A}}^{-1} \right)^{\frac{1}{2}} .
\end{equation}

Notice the similarity between (\ref{eq: E_DA pt}) and (\ref{eq: E_DA coral}), which coincide when the matrices $ \mathbf{\Sigma}_{\text{A}}$ and $ \mathbf{\Sigma}_{\text{S}}$ commute. 
So, the geometric perspective leads to a similar approach with a slightly different implementation. 
We remark that empirically this alternative based on PT with $\mE_{\text{PT}}$ leads to similar results as using the matrix $\mE$ in (\ref{eq: E_DA coral}).

\subsection{Theoretical Analysis}

For analysis purposes, we consider the \emph{population} correlation matrices,  neglecting the errors stemming from the finite sample size. For the population correlation matrices, we omit the hat symbol.
{The proofs for the theoretical results appear in the supplementary material. Note that in the experimental results in Section V, we do not consider the assumptions made in this section}

We start by examining the population correlation matrix of the adapted signal. 
\begin{proposition}
\label{prop: correlation alignment}
Assume the position of the {desired source and the interfering }sources are randomly distributed in their respective regions.

If
\begin{equation}
\label{eq: condition E[zz]=Sigma_P}
    \begin{split}
        \mathrm{E}[\vz(l)\vz^H(l) ] =  \mathbf{\Sigma}_{\text{A}},
    \end{split}
\end{equation}
Then
\begin{equation}
    \begin{split}
        \mathrm{E}[\vy(l)\vy^H(l)  ] =  \mathbf{\Sigma}_{\text{S}}        ,
    \end{split}
\end{equation}
%
\end{proposition}
According to Proposition \ref{prop: correlation alignment}, the proposed approach aligns the second-order statistics of the received signal from the operational domain with a signal generated in the reference domain.

In the following result, we provide an explicit expression for $\mathbf{\Sigma}_{\text{A}}$, the mean correlation matrix characterizing the operational environment, which offers insight into the proposed adaption.

\begin{proposition}
\label{prop: Sigma_S expressed using TFs}
    If the {desired source, the interfering} sources and the noise are all uncorrelated, then 
    \begin{equation}
             \mathbf{\Sigma}_{\text{A}} =                    
    \frac{1}{N_{\text{A}}}    
    \sum_{j=1}^{N_{\text{A}}}     
        \sigma_{s_j}^2 
         \vh_j\vh_j^H      + 
        \sum_{i=1}^{\tilde{N}_I}\sigma_{q_i}^2 \vg_i\vg_i^H + 
        \sigma_v^2\mI
    .   
    \end{equation}
    where 
    \[
        \sigma_{s_j}^2 =    \frac{1}{L}\sum_{l=1}^L |s_j(l)|^2
    \]
    is the signal power of the source at the $j$th position,
    \[
    \sigma_{q_i}^2 = \frac{1}{L}\sum_{l=1}^L |q_i(l)|^2
    \]
    is the signal power of the interfering source at the $i$th position, and $\tilde{N}_I$ is the total number of transmissions of interfering sources during the adaptation phase.

    

\end{proposition}
Before the array deployment, simulated steering vectors are generated to compute the matrix $\mathbf{\Sigma}_{\text{S}}$.
In contrast, during the adaptation phase, after the array deployment, we do not have direct access to the TFs characterizing the environment, and the received signals are used to compute the 
matrix $\mathbf{\Sigma}_{\text{A}}$.
So, the matrices $\mathbf{\Sigma}_{\text{S}}$ and $\mathbf{\Sigma}_{\text{A}}$ are computed using different entities. 
However, according to Proposition \ref{prop: Sigma_S expressed using TFs}, even though the matrix $\mathbf{\Sigma}_{\text{A}}$ is computed using the received signal, it comprises outer products of TFs. The proposed approach can be viewed as transporting the received signal from an operational environment with TFs that include interfering sources and indirect paths to the reference environment, considering solely the direct path.

In the rest of the analysis, we consider a statistical setting in which the positions of the different sources are 
 drawn independently from the same distribution.
 In this setting, all sources are located within the region of interest. The steering vector to the desired source, $\vd_S$, and the steering vector to the interfering source, $\vd_I$, are independent and identically distributed (i.i.d). 
This setting describes, for example, randomly generating a position for the interfering source and multiple positions for the desired source in each experiment and then repeating the experiment to examine the expected performance.

For the analysis, we make the following assumption regarding the population correlation matrix of the received signal.
 \begin{assumption}
     $\mathbf{\Sigma} =  \mathbf{\Sigma}_{\text{A}} + \vd_S\vd_S^H$ 
 \end{assumption}
The assumption implies that the environment remains unchanged in the adaptation and operational phases, except for the presence of the desired source whose DoA is estimated. 
This implies that only the interfering sources are active during the adaptation phase. In contrast, the desired sources are also active during the operational phase.

The following result shows that the quadratic term $|\vd^H(\theta)\mathbf{E}\vd(\theta)|^2$ with the steering vectors to the different sources is indicative of the spectrum of the DS beamformer.
\begin{proposition}
\label{prop: spectrum of E to spectrum EGammaE}
    If 
    \begin{equation}
    \label{eq: condition d_sEd_s > d_IEd_I}
    \begin{split}
        \mathbb{E}[|\vd_S^H\mE     
      \vd_S|^2] >
      \mathbb{E}[|\vd_I^H\mE     
      \vd_I|^2] ,
    \end{split}
\end{equation}
then
\begin{equation}
\label{eq: proposition 3 P_DS is better}
    \begin{split}
       \mathbb{E}[ P_{\text{DS}}(\mE\mathbf{\Sigma}\mE^H;\theta_S)]
        >
       \mathbb{E}[ P_{\text{DS}}(\mE\mathbf{\Sigma}\mE^H;\theta_I)].
    \end{split}
\end{equation}
\end{proposition}

According to Proposition \ref{prop: spectrum of E to spectrum EGammaE},
the interaction between the steering vectors and $\mE$ indicates the improvement obtained by the proposed approach.
In Section \ref{Sec: simulation results}, we present examples of the quadratic term $|\vd^H(\theta)\mathbf{E}\vd(\theta)|^2$ for different settings and show that condition (\ref{eq: condition d_sEd_s > d_IEd_I}) holds.

According to Proposition \ref{prop: correlation alignment}, the matrix $\mE$ can be viewed as derived by the objective of correlation alignment. Nonetheless, according to Proposition \ref{prop: spectrum of E to spectrum EGammaE}, it also leads to a meaningful term, $|\vd^H(\theta)\mathbf{E}\vd(\theta)|^2$.

 \subsection{Empirical Support}
\begin{figure}
\centering  
\subfloat[]{\includegraphics[width=0.48\linewidth]{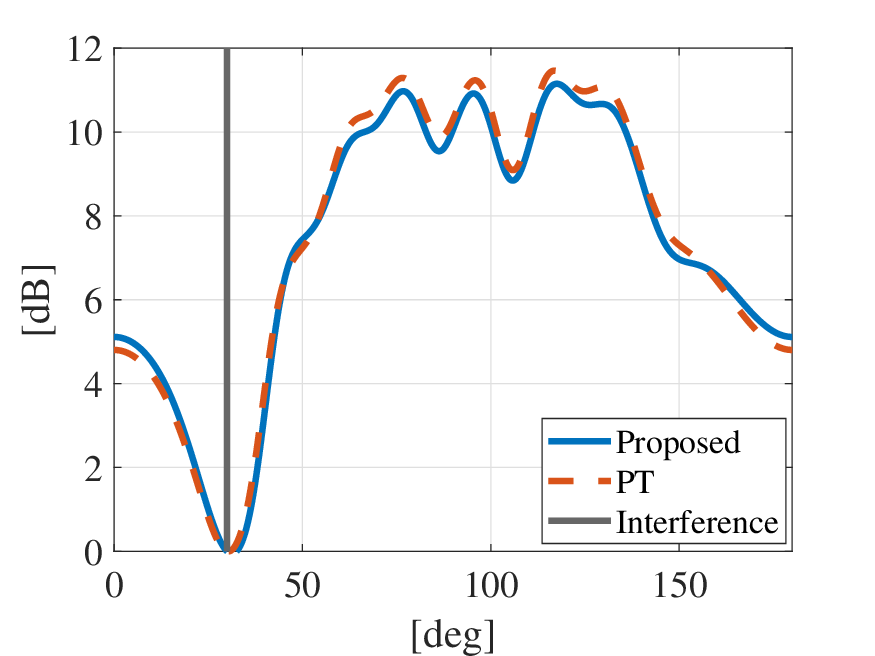}}
\subfloat[]{\includegraphics[width=0.48\linewidth]{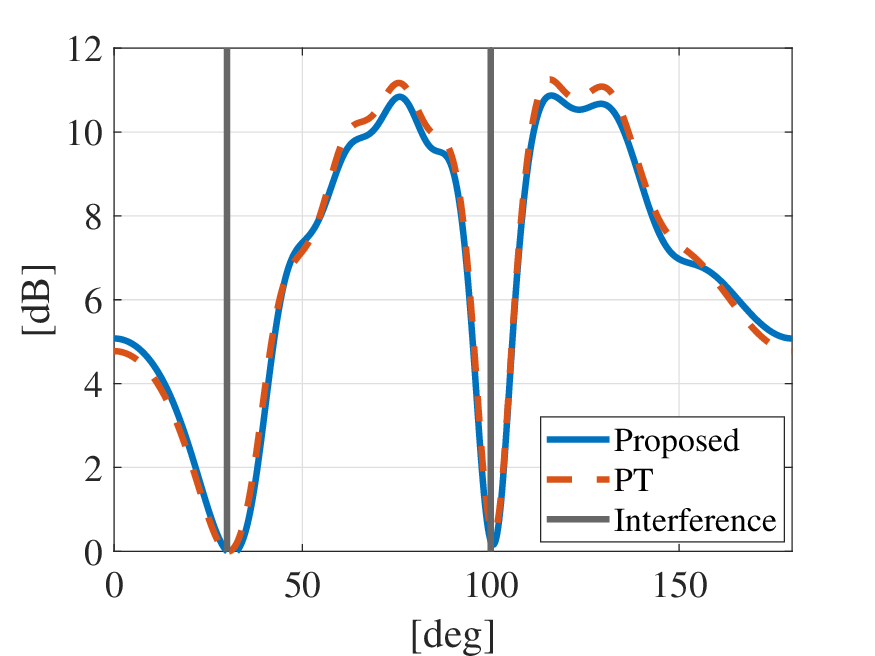}}
\caption{Theoretic expression for the output SIR in the presence of (a) A single interfering source. (b) Two interfering sources. The input SIR is $0\text{dB}$. The proposed approach using (\ref{eq: E_DA coral}) and (\ref{eq: E_DA pt}) appears in blue and red, respectively. The solid black line shows the direction of the interfering sources.
}
\label{fig: Theoretic SIR expression}
\end{figure}

 We present the following experiment, which examines the theoretical advantage of the proposed approach.
We consider desired and interfering sources with the same power positioned in an anechoic environment.
The population correlation matrix of the received signal is 
 \begin{equation}
     \mathbf{\Sigma} = 
     \sigma^2\vd_{S}\vd_{S}^H +
     \sigma^2\vd_I\vd_I^H +
     \sigma_v^2 \mI,
 \end{equation}
 where $\vd_{S}$ and $\vd_I$ are the steering vectors of the desired source and the interfering source, respectively, and $\sigma =  \sigma_{s_1} =  \sigma_{q_1}$.
 We define the output signal-to-interference ratio (SIR) associated with the correlation matrix $\mathbf{\Sigma}$ by
\begin{equation}
    \text{SIR} = \frac{P(\mathbf{\Sigma},\theta_S)}{P(\mathbf{\Sigma},\theta_I)},
\end{equation}
where $\theta_S$ and $\theta_I$ are the directions to the desired and interfering sources, respectively.
 So, the output SIR of the typically-used DS beamformer is 
 \begin{equation}
     \text{SIR}(\mathbf{\Sigma}) = 
     0\text{dB}.
 \end{equation}
 We simulatively examine the theoretic expression of the output SIR of the proposed approach with the DS beamformer for a phased array of $M=9$ elements, i.e.,
  \begin{equation}
     \text{SIR}(\mE\mathbf{\Sigma}\mE^H) = 
     \frac
     {P_{\text{DS}}(\mE\mathbf{\Sigma}\mE^H,\theta_{S})}
     {P_{\text{DS}}(\mE\mathbf{\Sigma}\mE^H,\theta_I)}    ,
 \end{equation}
 for different directions of the desired source, $\theta_{S}$. 
 The direction of the interfering source is set to $\theta_I=30\degree$.
 For the computation of the matrix $\mE$, we consider $N_{\text{A}} = 100$ different positions during the adaptation phase and $N_{\text{S}} = 100$ simulated steering vectors.
 Figure \ref{fig: Theoretic SIR expression}(a) presents the results for the proposed approach as well as the results for $\mE_{\text{PT}}$ in (\ref{eq: E_DA pt}). We see that for all the directions of the desired source, the output SIR is greater than $0\text{dB}$, meaning the spectrum of the DS beamformer with the proposed approach reduces the interfering source. Estimating the DoA by the argmax of the spectrum according to (\ref{eq: theta hat as argmax}) results in an accurate estimation of the desired source rather than estimating the DoA of the interfering source instead.
 In addition, the results for the map using $\mE$ in (\ref{eq: E_DA coral}) and $\mE_{\text{PT}}$ in (\ref{eq: E_DA pt}) are very similar.
Figure \ref{fig: Theoretic SIR expression} (b) is the same as  Figure \ref{fig: Theoretic SIR expression} (a) only for two interfering sources. Even with two simultaneously active interfering sources, the proposed approach leads to higher SIR values than the commonly used DS beamformer.

 \section{Simulation Results}
\label{Sec: simulation results}

In this section, we showcase the proposed approach's advantages in two scenarios: a reverberant acoustic environment and an RF setting\footnote{The code is available at  \href{https://github.com/amitaybar/Domain-Adaptation-for-DoA-Estimation-in-Multipath-Channels-with-Interferences}{https://github.com/amitaybar/Domain-Adaptation-for-DoA-Estimation-in-Multipath-Channels-with-Interferences}}.
In both cases, the desired source is accompanied by interfering sources. 
{The proposed approach serves a data enhancement and is applied before a subsequent DoA estimation method. Therefore,}
to illustrate the versatility of our approach, we examine its performance using three widely used beamformers: the DS, the MVDR, and MUSIC, which represent three different approaches for DoA estimation. The proposed DA could also be applied to other DoA estimation methods.
For quantitative evaluation, we use the absolute value of the DoA estimation error given by
\begin{equation}
    \Delta \theta_i = | \hat{\theta}_i-\theta_i |,
\end{equation}
where $\theta_i$ is the DoA of the $i$th source, and  $\hat{\theta}_i$ is the DoA estimation of the $i$th source computed using the different beamformers  (\ref{eq: theta hat as argmax}).

\subsection{The Acoustic Setting}

\begin{figure}
  \begin{center}  
\includegraphics[width=0.75\linewidth]{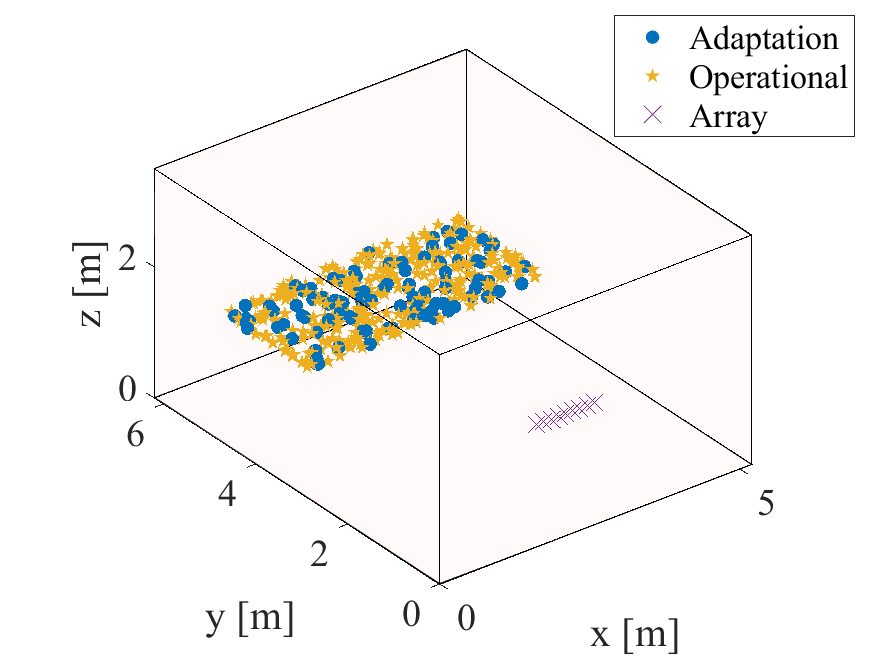}
\caption{A $3\text{D}$ view of the acoustic room. The positions of the desired source in the adaptation and operational phases are marked by blue circles and orange asterisks, respectively. A purple ‘x’ marks the phased array.
}
\label{fig: Acoustic Room}
\end{center}
\end{figure}

We consider a microphone array of $M=9$ microphones in a reverberant room of dimensions $5.2\text{m} \times 6.2\text{m} \times 3.5\text{m}$. The reference microphone is positioned at $(2\text{m}, \; 0.5\text{m}, \;1.5\text{m})$ and the rest of the microphones are distant $12\text{cm}$ apart along the X-axis.  
In all the experiments, the region of interest is a rectangular of dimensions $4\text{m}\times2\text{m}$ in the XY plane with two corners at $(0.5\text{m},\;  3.5\text{m},\;  1.5\text{m})$ and $(0.5\text{m},\;  5.5\text{m},\;  1.5\text{m})$, and the other two corners are distant $4\text{m}$ along the X-axis.
Figure \ref{fig: Acoustic Room} presents the room. A purple ‘x’ marks the position of a microphone, a blue circle marks the position of a source during the adaptation phase, and an orange asterisk denotes the position of a source during the operational phase.
The positions of the simulated steering vectors, generated before the array's deployment, are omitted for presentation purposes.

The signal of each source is generated as Gaussian noise using the standard normal distribution. Its duration is $2.5\text{s}$, sampled at $12\text{KHz}$ resulting in $30,000$ samples. {The received signal is processed in the time-frequency domain using the short-time Fourier transform (STFT) }with a rectangular analysis window of $2048$ samples and $50\%$ overlap. We focus on a single frequency bin (indexed $242$ and corresponds to a frequency of approximately $1.42\text{KHz}$), so the distance between the microphones is half the wavelength. 
The acoustic impulse response (AIR) between the different sources and each microphone is modeled using the image method \cite{allen1979image} as implemented by the room impulse response simulator in \cite{habets2006room}. Each AIR consists of $2048$ samples.
In all the experiments, the SNR at each microphone is $20\text{dB}$.
{We note that further improvement in the DoA estimation can be made by fusing multiple frequency bins. We demonstrate the proposed approach on a single-frequency bin and leave the fusion for future work. }

In the first experiment, we examine the proposed approach's performance for different reverberation times. The number of varying transmission positions in the adaptation phase is $100$, which leads to $100$ received signals. These signals could originate from several sources from different places or from a single source from $100$ different positions. 
We note that for a signal duration of $2.5\text{s}$, the total duration of all the signals at the adaptation phase is slightly above $4$ minutes.
The number of reference signals generated in the pre-deployment phase is $200$.

\begin{figure}
\centering
\subfloat[]{\includegraphics[width=0.5\linewidth]{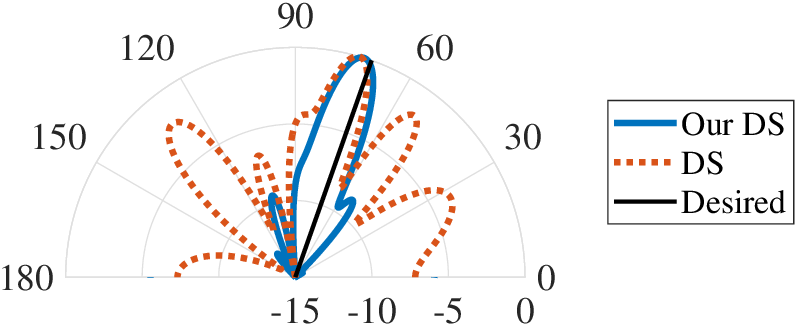}}
\subfloat[]{\includegraphics[width=0.5\linewidth]{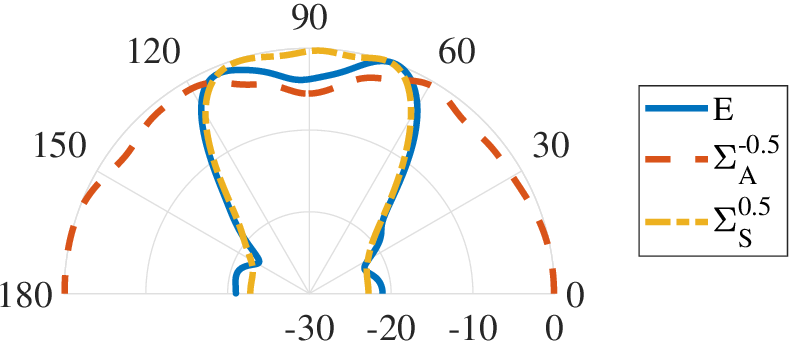}}
\caption{(a) The spectrum of the proposed approach with the DS beamformer (blue) and the typically used DS beamformer (red) in the reverberant room. The solid black line shows the direction of the desired source.
(b): The quatratic terms $|\vd^H(\theta)\mE\vd(\theta)|^2$, $|\vd^H(\theta)\mathbf{\Sigma}_{\text{A}}^{-\frac{1}{2}}\vd(\theta)|^2$, and $|\vd^H(\theta)\mathbf{\Sigma}_{\text{S}}^{\frac{1}{2}}\vd(\theta)|^2$ in blue, red, and orange, respectively.}
\label{fig: Spectrum Only revs acoust} 
\end{figure}

We start with an example of the spectrum of the DS beamformer in the reverberant room. Figure \ref{fig: Spectrum Only revs acoust} (a) presents the spectrum of the proposed approach with the DS beamformer and the spectrum of the commonly-used DS beamformer in blue and red, respectively, for a reverberation time of $\beta=400\text{ms}$. The black solid line indicates the direction of the desired source. 
Both spectra present main lobes that coincide with the direction of the desired source, albeit with a minor bias (the main lobe of the proposed approach is slightly closer).
Other nuisance lobes exist for the widely used DS beamformer due to reverberations. In contrast, the proposed approach results in a main lobe,
while the other lobes are significantly attenuated. 

Figure \ref{fig: Spectrum Only revs acoust} (b) presents the absolute value of the quadratic term of the matrix $\mE$, i.e., $|\vd^H(\theta)\mE\vd(\theta)|^2$, in blue. We refer to it as the spectrum of $\mE$. We note that the section bounding the upper line of the rectangular region of interest is $\theta \in [68\degree,112\degree]$, and we see that the spectrum of $\mE$ is directed to that section. We recall that condition (\ref{eq: condition d_sEd_s > d_IEd_I}) in proposition \ref{prop: spectrum of E to spectrum EGammaE} involved the quadratic term of $\mE$. 
By considering the reverberations as interfering sources, we observe that the quadratic term takes on higher values for sources located within or paths originating from the region of interest.
 The matrix $\mE$ is a product of the two matrices, $ \mathbf{\Sigma}_{\text{{A}}}^{{-\frac{1}{2}}}$ and $ \mathbf{\Sigma}_{\text{S}}^{\frac{1}{2}}$, whose spectra using the DS beamformer appear in red and orange, respectively. We see that the spectrum of $ \mathbf{\Sigma}_{\text{{A}}}^{{-\frac{1}{2}}}$ is close to omnidirectional and the spectrum of $ \mathbf{\Sigma}_{\text{S}}^{\frac{1}{2}}$ is similar to the spectrum of $\mE$.  Even though these spectra look similar, using $\mE$ still leads to better results. The reason is that the spectrum only considers the absolute value, whereas the vector's direction is also significant. 
 This result sheds some light on the effect of $\mathbf{E}$ on the adapted signal and the resulting beamforming spectrum in focusing the beam on the region of interest, which is "learned" implicitly from the simulated steering vectors and captured by the matrix $\mathbf{E}$.
 
Next, we examine the performance of the proposed approach for different reverberation times. 
We evaluate the DoA estimation using $300$ randomly generated source positions.
\begin{figure}
\centering
\subfloat[]{\includegraphics[width=0.80\linewidth]{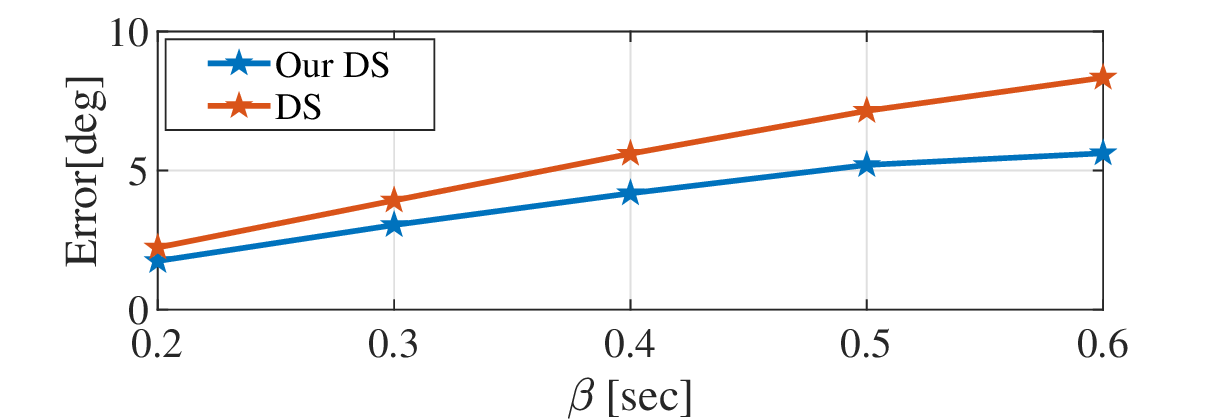}}\\
\subfloat[]{\includegraphics[width=0.80\linewidth]{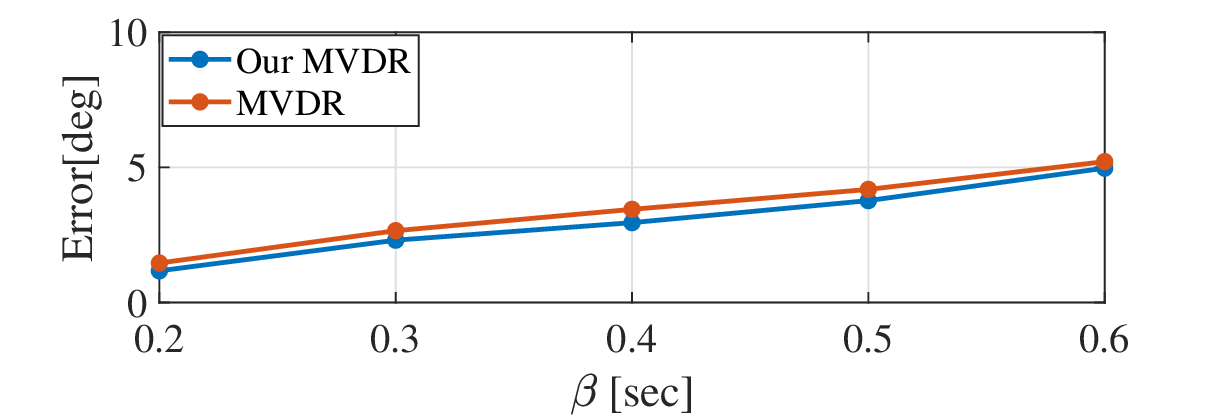}}\\
\subfloat[]{\includegraphics[width=0.80\linewidth]{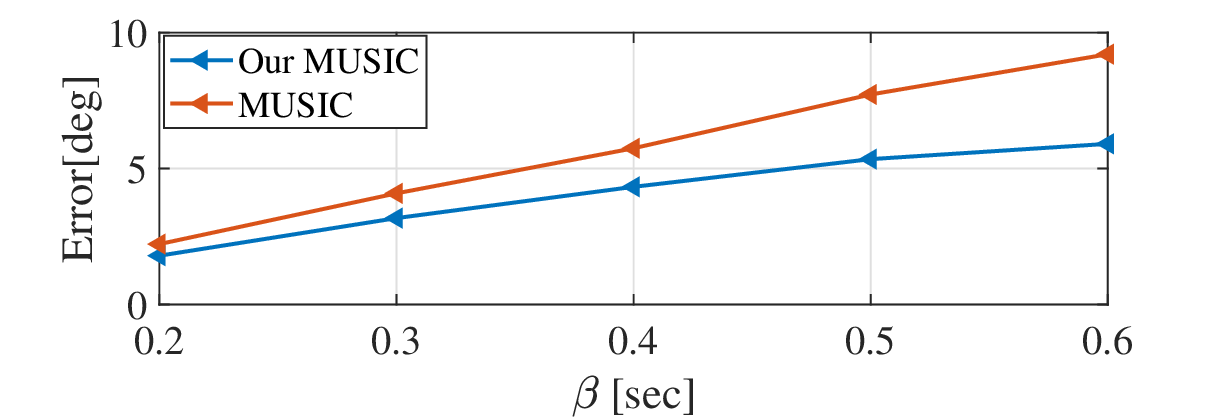}}
\caption{The median of the DoA estimation errors for different reverberation time values $\beta$ for (a) the DS beamformer, (b) the MVDR, and (c) MUSIC.
}
\label{fig: DoA for different beta}
\end{figure}
Figure \ref{fig: DoA for different beta} presents the obtained DoA estimation error as the function of the reverberation time for the DS beamformer, the MVDR, and MUSIC.
For the MVDR beamformer, we empirically found that to adapt to a reverberant environment, it is best to adapt the inverse of the sample correlation matrix of the received signal, $\sSigma^{-1}$. We follow the same steps as in Algorithm \ref{Alg: DA for DoA}, only with the following set of matrices $\big\{\left( \mathbf{\Sigma}_j^{\text{S}}\right)^{-1}\big\}_{j=1}^{N_{\text{S}}}$ and $\big\{( \sSigma_i^{\text{A}})^{-1}\big\}_{i=1}^{N_{\text{A}}}$. 
The figure shows that the longer the reverberation time is, the larger the DoA estimation errors are. Importantly, for all the tested reverberation times and all the tested beamformers, the proposed approach leads to improved results, up to $\sim2[\text{deg}]$ improvement for $\beta=500\text{ms}$. Furthermore, the longer the reverberation time is, the greater the improvement of the proposed approach is. Following Proposition \ref{prop: Sigma_S expressed using TFs}, we can view the proposed approach as transporting the received signal from the operational domain of the reverberant environment to the reference domain of an anechoic environment. The reverberations can be viewed as the domain shift between the two domains. We see that the larger the domain shift is, the more significant the improvement of the proposed approach becomes.

\begin{figure}
  \begin{center}
\includegraphics[width=0.85\linewidth]{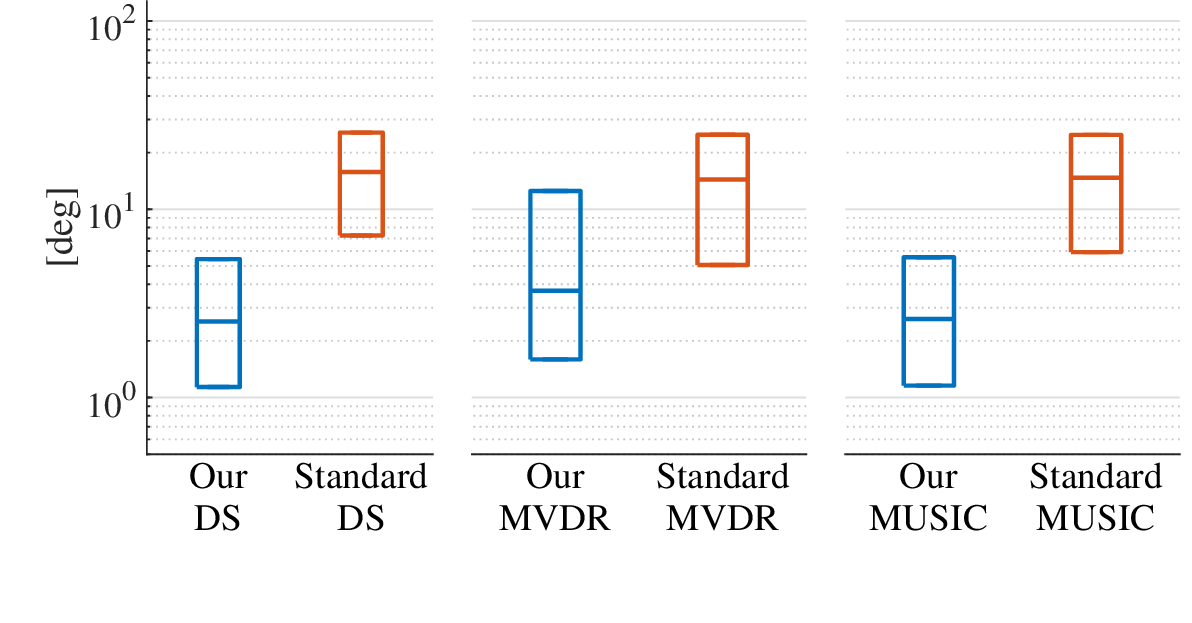}
\caption{DoA estimation errors in the presence of an interfering source with SIR $-20\text{dB}$ for the proposed approach and the standard beamformers in blue and red, respectively.}
\label{fig: DoA performance acoust no interference}
\end{center}
\end{figure}

In the next experiment, we consider an interfering source positioned uniformly at random within the region of interest. 
The SIR is $-20\text{dB}$, i.e., the interfering source is much stronger than the desired source.
The SIR is defined with respect to the transmitted signals.
The number of signals in the adaptation phase is $100$, and the number of reference signals is $200$. The number of positions in the operational phase is $200$. We repeat this experiment for $20$ different interfering source positions. All sources' positions are generated uniformly at random in the region of interest. 
Figure \ref{fig: DoA performance acoust no interference} presents the DoA estimation errors of the DS, the MVDR, and the MUSIC beamformers, obtained by the proposed approach and the baseline approach, in blue and red, respectively. The box indicates the $25$th and the $75$th percentiles, and the horizontal line indicates the median.

We see that the proposed approach improves the performance of all the beamformers by more than $10\degree$.
We recall that the proposed approach with MUSIC considers a signal space of a single dimension even in the presence of the interfering source. The fact that the proposed method with MUSIC results in improved performance indicates that the proposed approach leads to the rejection of the interfering source, maintaining the desired source as the dominant eigenvector of the correlation matrix.
The standard beamformers estimate the DoA of the interfering sources, which results in large errors.

We repeat the experiment for different SIR values. 
Table \ref{Table: Acoustic one interference} presents the median of the DoA estimation error, along with the interquartile range (IQR) in parenthesis, for the proposed approach and the typically-used beamformers.
We see a significant improvement in the accuracy of the DoA estimation when the proposed approach is applied to all the tested beamformers. %
In addition, the proposed approach results in smaller IQR values. 

\begin{table*}[t]
\caption{The median and IQR (in parenthesis) of the DoA estimation error for the acoustic setting.}
\centering
\begin{tabular}{{c||c|c||c|c||c|c|}}
\multicolumn{1}{c||}{\bf Beamformer} 
&\multicolumn{1}{c|}{\bf DS}
&\multicolumn{1}{c||}{\bf Our DS }
&\multicolumn{1}{c|}{\bf MVDR}
&\multicolumn{1}{c||}{\bf Our MVDR}
&\multicolumn{1}{c|}{\bf MUSIC}
&\multicolumn{1}{c|}{\bf Our MUSIC}
 
 \\

\hline 
  SIR=$-20\text{dB}$   & 14.7 (16.0)     & \textbf{2.6} (4.3)     & 13.9 (17.6)     & \textbf{3.6} (9.0)     & 14.5 (17.3)     & \textbf{2.6} (4.2)  \\

 \hline 
  SIR=$-10\text{dB}$   & 14.5 (16.4)     & \textbf{1.9} (2.9)     & 10.3 (17.9)     & \textbf{2.6} (5.1)     & 6.0 (16.2)     & \textbf{2.0} (2.9)  \\

 \hline 
  SIR=$0\text{dB}$   & 4.5 (12.7)     & \textbf{1.8} (2.6)     & 4.1 (14.9)     & \textbf{2.1} (3.7)     & 5.2 (15.1)     & \textbf{1.8} (2.6)  \\
  



\end{tabular}
\label{Table: Acoustic one interference}
\end{table*}

In the next experiment, the interfering source is positioned in a rectangular region of $1\text{m}\times 2\text{m}$, separate from the region of interest. The farthest corner of the interference region from the array is $(0\text{m}, \; 3\text{m}, \; 1.5\text{m})$. 
In the adaptation phase, a source is active at $100$ different positions, chosen uniformly at random within the region of interest. For each such position, an interfering source positioned uniformly at random within the interference region is also active. The number of reference signals is $200$. During the operational phase, the desired source is positioned uniformly at random in the region of interest, and \emph{two} interfering sources are placed uniformly at random in the interference region.
The SIR is $0\text{dB}$ and
the reverberation time is $\beta = 200\text{ms}$. The results are evaluated on $300$ different positions. 
We emphasize that the interference region remains unknown during the adaptation and operational phases. Moreover, the relative position between the region of interest and the interference region is also unknown.

\begin{figure}
  \centering
\subfloat[]{\includegraphics[width=0.65\linewidth]{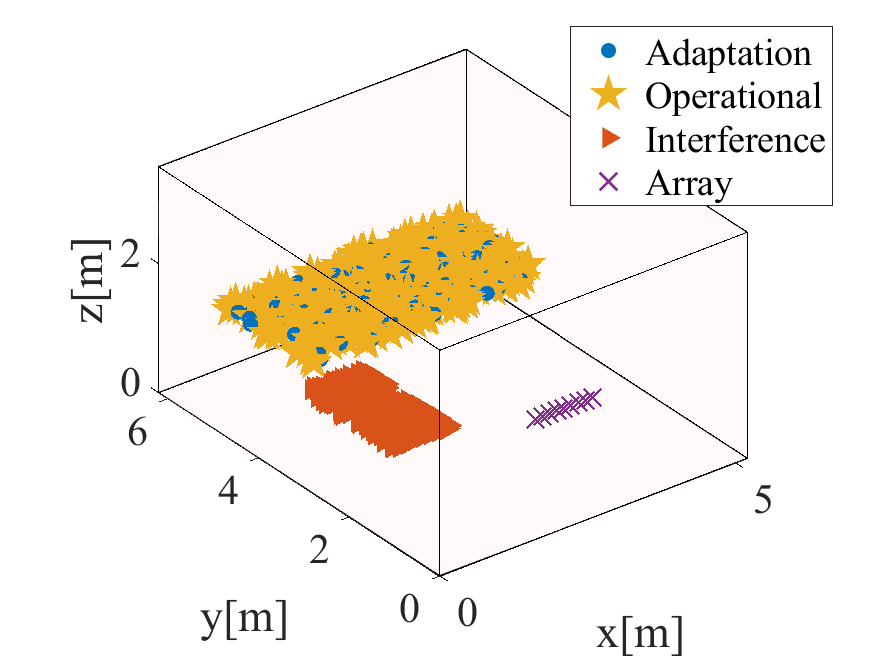}}
\\
\subfloat[]{\includegraphics[width=0.75\linewidth]{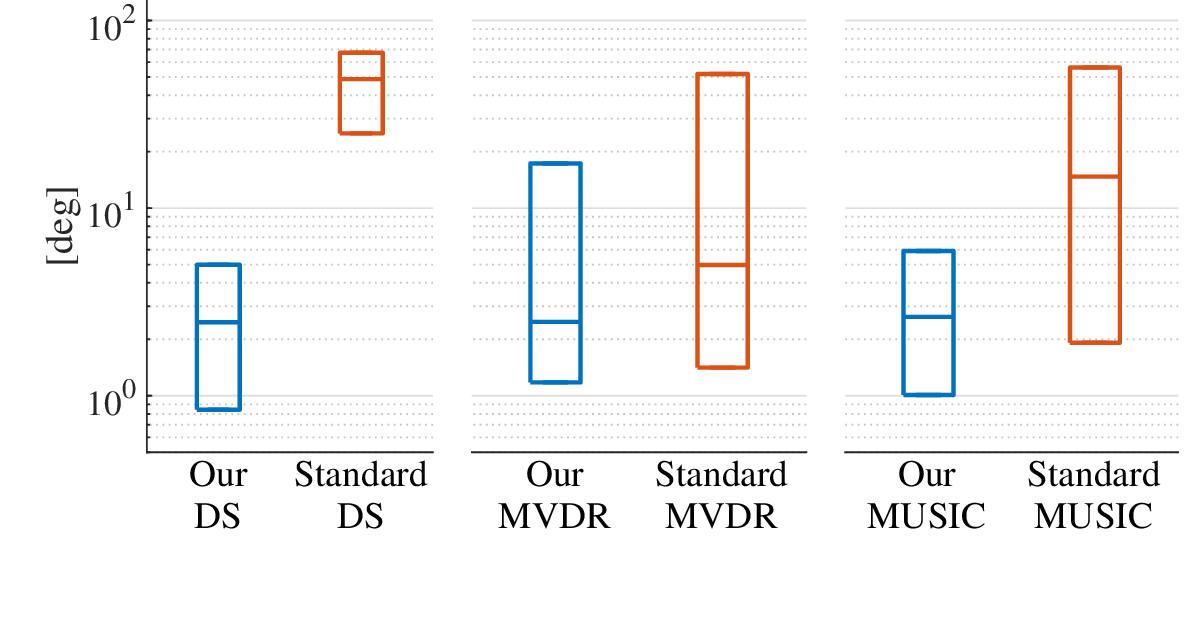}}
\caption{
(a) A $3\text{D}$ view of the room with the interference region. The positions of the desired source in the adaptation and operational phases are marked by blue circles and orange asterisks, respectively. The interfering source appears in a red triangle. A purple ‘x’ marks the phased array. (b) The DoA estimation errors in the presence of two simultaneously active interfering sources with SIR $0\text{dB}$.}
\label{fig: Room with interference section and DoA perfrmance} 
\end{figure}

Figure \ref{fig: Room with interference section and DoA perfrmance}(a) is the same as Figure \ref{fig: Acoustic Room}, presenting the room with the positions of the different sources of this scenario.
The positions of the interfering source appear in red triangles.
Figure \ref{fig: Room with interference section and DoA perfrmance}(b) presents the DoA estimation errors obtained by the DS, the MVDR, and the MUSIC beamformers when using the proposed approach and the baseline, in blue and red, respectively. 
For all the beamformers, there is a clear improvement in the DoA estimations using the proposed approach. 

We emphasize that the proposed approach performs well in the presence of two interfering sources, although the adaptation phase consisted of only one active interfering source at a time. 
In the adaptation phase, the interference region is learned, and the proposed approach can be viewed as leading to a null section towards the interference region. 
Furthermore, if the interference region is a priori known, steering vectors from the interference region can be generated before the deployment instead of collecting signals in the adaption phase, making the adaptation phase redundant. As a result, the proposed approach can also be viewed as a method of implicitly designing beamformers with null steering towards known interference regions.

\subsection{The RF Setting}

\begin{figure}[t]
  \begin{center}
\includegraphics[width=0.55\linewidth]{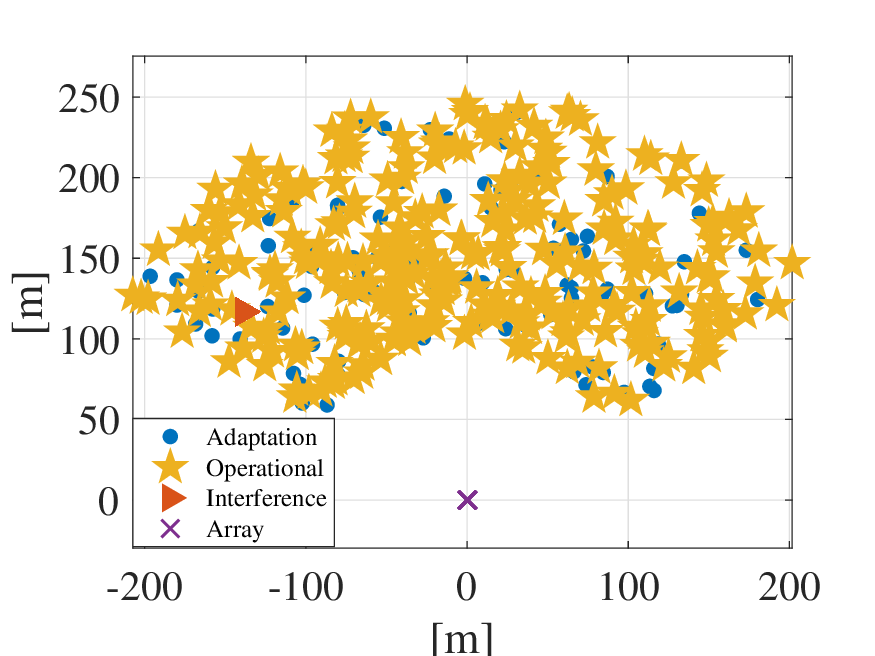}
\caption{The RF scenario. The positions of the desired source in the adaptation and operational phases are marked by blue circles and orange asterisks, respectively. The interfering source appears in a red triangle. A purple ‘x’ marks the phased array.}
\label{fig: RF Room}
\end{center}
\end{figure}

We consider a narrowband signal at the frequency of $2.4\text{GHz}$. The phased array consists of $9$ elements distant $6.25\text{cm}$ apart along the x-axis, where the reference element is positioned at the origin. The region of interest is a section of $120\degree$ between $30\degree$ and $150\degree$ and a radius between $100\text{m}$ and $250\text{m}$. Figure \ref{fig: RF Room} presents the scenario with the different sources. It is the same as Figure \ref{fig: Acoustic Room} only for the RF setting and a top view.

The signal from all the sources is simulated as complex Gaussian noise. 
We consider a channel comprising the direct path as follows. 
The TF of the channel between the source and the $m$th element is given by  \cite{krim1996two}
\begin{equation}
    h_m = \frac{a}{r_m}e^{-j2\pi \frac{r_m}{\lambda}},
\end{equation}
where $a$ is a constant and $r_m$ is the distance between a source and the $m$th element of the array. We set $a=1$ without loss of generality since the results are presented as a function of the SNR. 
%
%
In all the experiments, we consider $100$ different positions in the adaptation phase and $200$ reference signals.

\begin{figure}
\centering
\subfloat[]{\includegraphics[width=0.5\linewidth]{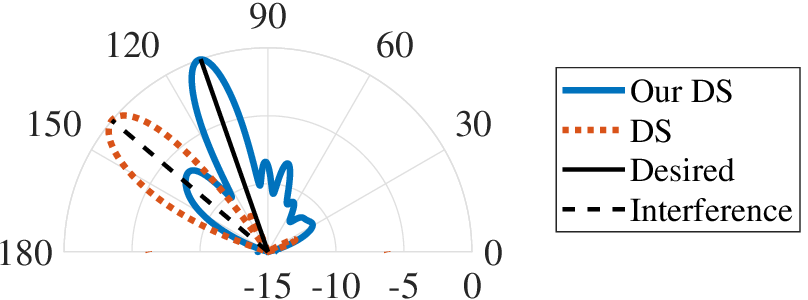}}
\subfloat[]{\includegraphics[width=0.5\linewidth]{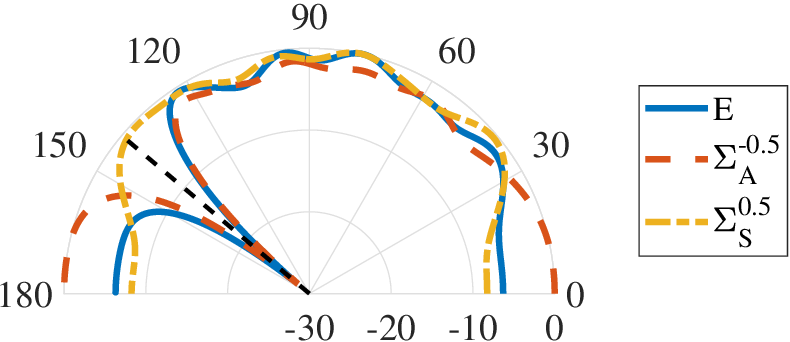}}
\caption{(a) The spectrum of the proposed approach with the DS beamformer (blue) and the typically used DS beamformer (red) in the presence of an interfering source with SIR $-20\text{dB}$.
(b) The quadratic terms of $|\vd^H(\theta)\mE\vd(\theta)|^2$, $|\vd^H(\theta)\mathbf{\Sigma}_{\text{A}}^{-\frac{1}{2}}\vd(\theta)|^2$, and $|\vd^H(\theta)\mathbf{\Sigma}_{\text{S}}^{\frac{1}{2}}\vd(\theta)|^2$ in blue, red, and orange, respectively, in the presence of an interfering source with SIR $-20\text{dB}$.
}
\label{fig: DS spectrum RF}
\end{figure}

In Figure \ref{fig: DS spectrum RF}, we present an example of the spectrum of the DS beamformer in the presence of an interfering source. Figure \ref{fig: DS spectrum RF}(a) presents the spectrum for the proposed approach with the DS beamformer and the typically-used beamformer in blue and red, respectively, for SIR $-20\text{dB}$. The black solid line shows the direction of the desired source, and the black dashed line indicates the direction of the interfering source. We see that the DS spectrum of the proposed approach results in a main lobe directed at the desired source. In contrast, the typically-used DS spectrum points toward the interfering source. 
Figure \ref{fig: DS spectrum RF}(b) presents the 
quadratic terms
$|\vd^H(\theta)\mE\vd(\theta)|^2$, $|\vd^H(\theta)\mathbf{\Sigma}_{\text{A}}^{-\frac{1}{2}}\vd(\theta)|^2$, and $|\vd^H(\theta)\mathbf{\Sigma}_{\text{S}}^{\frac{1}{2}}\vd(\theta)|^2$, in blue, red, and orange, respectively. The dashed line shows the direction of the interfering source. We see that the spectrum of $\mE$ results in null steering in the direction of the interfering source in addition to attenuation at the directions outside the region of interest. 
In light of Proposition \ref{prop: spectrum of E to spectrum EGammaE}, this leads to a higher output SIR for the proposed approach, as demonstrated in Figure \ref{fig: DS spectrum RF}(a).
Additionally, we see that the spectrum of the matrix $ \mathbf{\Sigma}_{\text{A}}^{{-\frac{1}{2}}}$ has a null towards the interfering source. However, it also leads to high values in directions outside the region of interest. The spectrum of the matrix $ \mathbf{\Sigma}_{\text{S}}^{\frac{1}{2}}$ attains high values at the region of interest but also at the direction of the interfering source.

\begin{figure}
\centering
\subfloat[]{\includegraphics[width=0.5\linewidth]{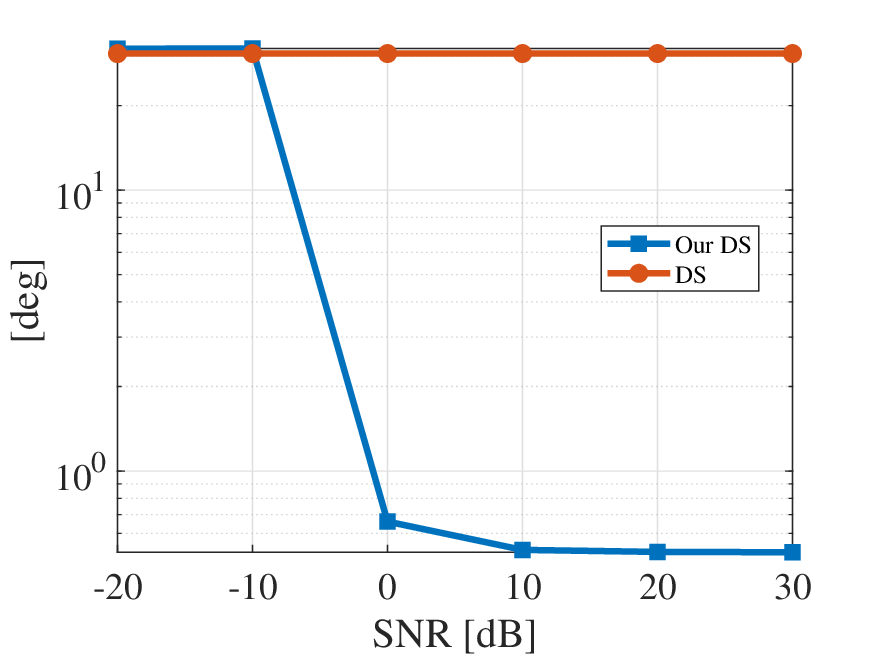}}
\subfloat[]{\includegraphics[width=0.5\linewidth]{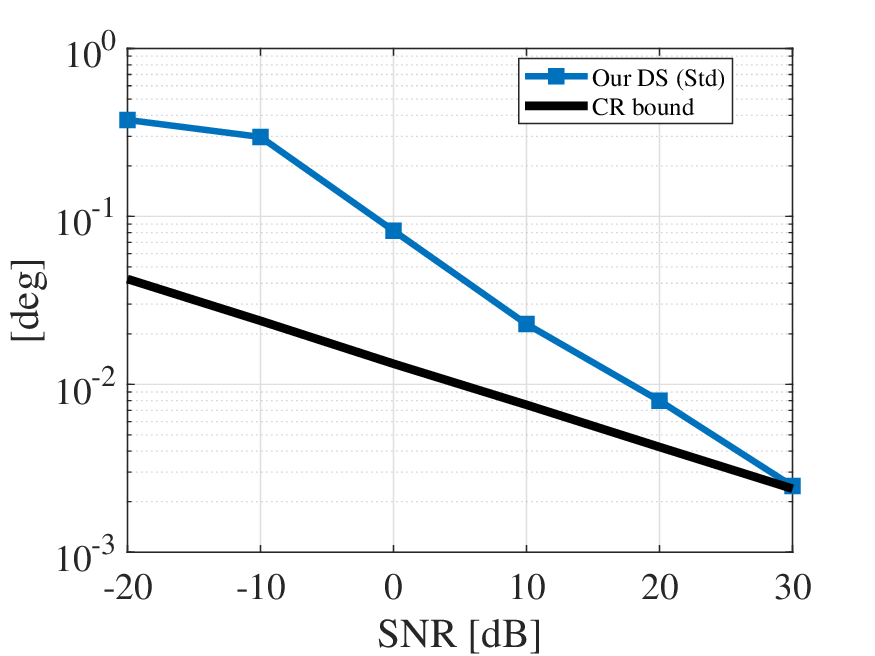}}
\caption{{(a) The RMSE of the DoA estimations of the proposed approach with the DS beamformer (blue) and the typically used DS beamformer (red) in the presence of an interfering source with SIR $-20\text{dB}$.
(b) The std of the proposed approach (blue) and the CRB (black).}
}
\label{fig: RMSE AND CRB}
\end{figure}

{
We continue by examining the performance of the proposed approach for different SNR values (considering the fixed mapping matrix $\mE$). Figure \ref{fig: RMSE AND CRB}a presents the RMSE of the proposed approach applied to the DS beamformer and the typically used DS beamformer. The typically used DS results in a high RMSE value fixed across different SNRs. This results from its spectrum pointing in the direction of the interfering source. In contrast, the proposed approach reduces the effect of the interfering source, allowing for an accurate DoA estimation of the desired source. For SNR values higher than $10\text{dB}$, the performance of the proposed approach is approximately fixed around 0.5 $[\text{deg}]$. This is due to a bias in the DoA estimations (note the bias of the DS is $\sim 30[\text{deg}]$). 
Next, we compare the performance of the proposed approach to the Cram\'er-Rao bound (CRB). The CRB assumes an unbiased estimator, so we focus on the standard deviation (std) instead of the RMSE. Note that the bound assumes a single desired source without the interfering source (we consider a CRB that does not address the association of the peaks in the spectrum with different sources). Figure \ref{fig: RMSE AND CRB}b presents the std of the proposed approach in blue and the CRB in black. We see that the higher the SNR is, the closer the std of the proposed approach to the CRB. 
}

\begin{figure}
  \begin{center}
\includegraphics[width=0.75\linewidth]{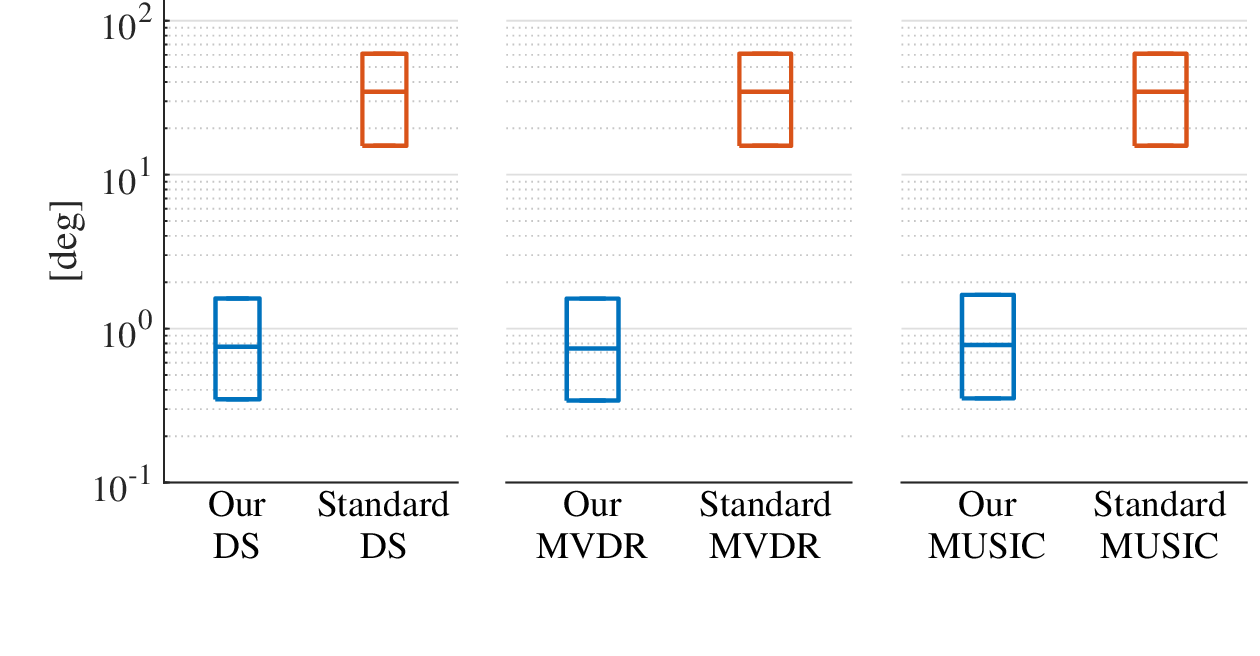}
\caption{The DoA estimation errors in the presence of an interfering source with SIR $-20\text{dB}$ for the proposed approach (blue) and the standard beamformers (red) in the log scale.}
\label{fig: DoA performance for RF single interfering source}
\end{center}
\end{figure}

In the next experiment, we examine the DoA estimation performance in the presence of a constantly active interfering source positioned at the region of interest. The SIR is set to $-20\text{dB}$, and the SNR is $20\text{dB}$. We evaluate the performance using $300$ different desired source positions. The positions of the desired source and the interfering source are generated uniformly at random in the region of interest. We repeat the experiment for $20$ different interfering source positions. Figure \ref{fig: DoA performance for RF single interfering source} shows the DoA estimation errors for the DS, MVDR, and MUSIC beamformers similarly to Figure \ref{fig: DoA performance acoust no interference}.
The proposed approach with MUSIC considers a signal space of a single dimension even in the presence of interfering sources. For the standard MUSIC, we assume to know the number of active sources.
We see that the proposed approach improves accuracy for all the beamformers.
The standard beamformers lead to large errors since they result in the estimation of the DoA of the interfering source, which is stronger than the desired source.
We repeat the experiment for different SNR and SIR values; the results are similar. See more detail in the SM. %

\begin{figure}[ht]
  \begin{center}
\includegraphics[width=0.6\linewidth]{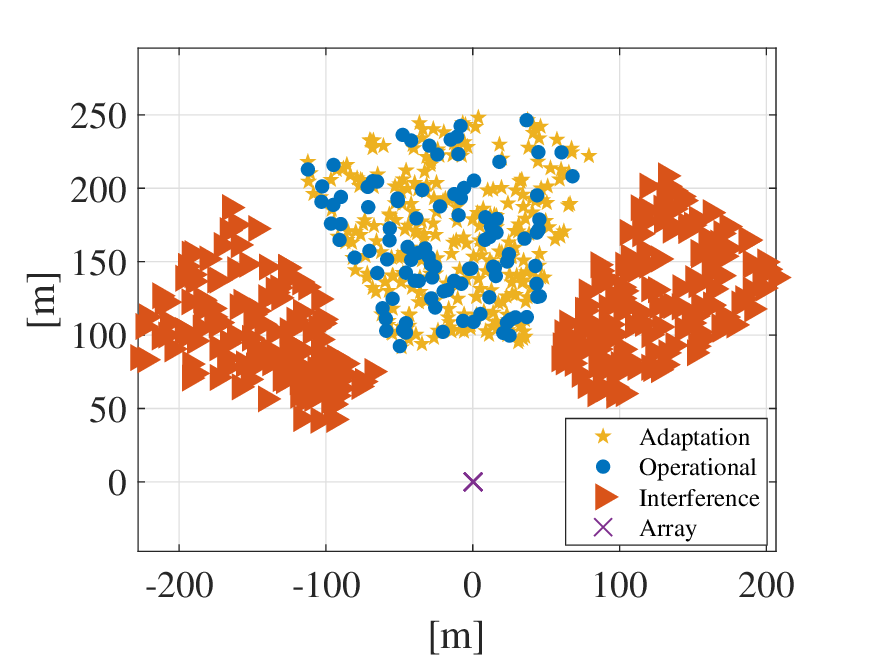}
\caption{The RF setting with a desired source section, and two disjoint interfering source sections. The positions of the desired source in the adaptation and operational phases are marked by blue circles and orange asterisks, respectively. The interfering sources appear in red triangles. A purple ‘x’ marks the phased array. }
\label{fig: RF Room interference section}
\end{center}
\end{figure}

We conclude with the following setting, demonstrating that the proposed approach can be applied to a scenario where the interference region consists of two separate regions. The region of interest is set between $70\degree$ and $120\degree$, and the interference region is the two separate sections $\theta \in [30\degree,60\degree] \cup [130\degree,160\degree]$. The radii remain the same. 
In the adaptation phase, each transmission of a desired source is contaminated by an interfering source. 
We evaluate the performance using $500$ different sources.
The positions of the different sources are generated uniformly at random in their respective regions.
The SNR is $20\text{dB}$.
Figure \ref{fig: RF Room interference section} is the same as Figure \ref{fig: RF Room}, presenting the setup of the current scenario.

\begin{figure}[ht]
  \centering
\subfloat[]{\includegraphics[width=0.75\linewidth]{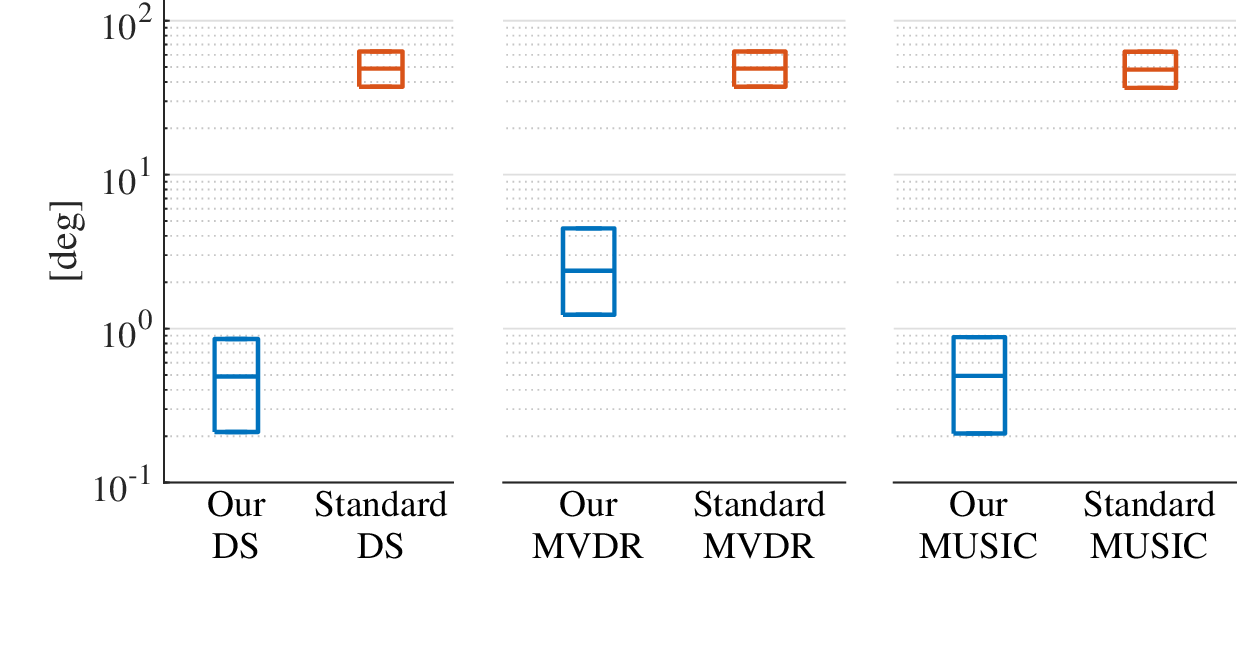}}
\\
\subfloat[]{\includegraphics[width=0.75\linewidth]{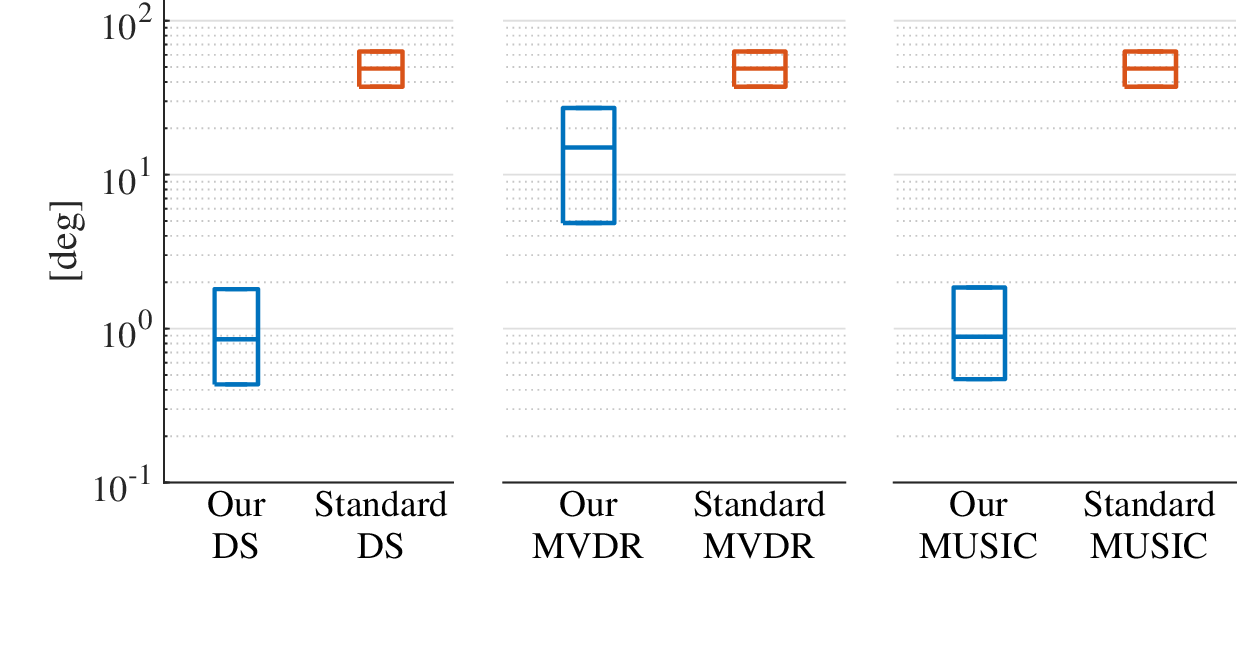}}
\caption{The DoA estimation errors in the presence of an interfering source for the proposed approach (blue) and the standard beamformers (red) with (a) SIR of $-10\text{dB}$ and (b)
SIR of $-20\text{dB}$.}
\label{fig: RF Room interference section and DoA performance} 
\end{figure}

Figure \ref{fig: RF Room interference section and DoA performance} presents the performance of the DoA estimation for the proposed approach and the typically used beamformers in blue and red, respectively. It is the same as Figure \ref{fig: DoA performance for RF single interfering source} only for the current setting and for SIR $-10\text{dB}$ (a) and for SIR $-20\text{dB}$ (b). We see that for all the tested beamformers, the proposed approach leads to better DoA estimation accuracy.
The standard beamformers estimate the DoA of the interfering sources, which results in large errors.
This experiment is an example of using the proposed approach for null section to separate sections,
even when the sections are unknown apriori. 

\begin{figure}[ht]
  \begin{center}
\includegraphics[width=0.6\linewidth]{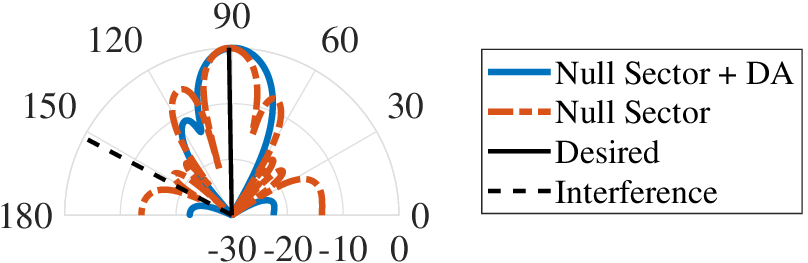}
\caption{{The beam pattern of the null section method in \cite{amar2015linearly} after applying the DA mapping to the sample correlation matrix (blue) and the beam pattern of the null section method in \cite{amar2015linearly} (red).}}
\label{fig: Null section + DA}
\end{center}
\end{figure}

{Next, we consider the work in \cite{amar2015linearly}, which proposed a method for designing a beam pattern with null to a known section (null section). For our version of the algorithm, we follow \cite{amar2015linearly} with one modification: we utilize the sample covariance matrix \textbf{after applying the map $\mE$}. Figure \ref{fig: Null section + DA} shows the beam pattern of the null section method (red) and the null section method using the mapped sample covariance matrix (blue). By applying the map, we see that the interfering source is further attenuated. This demonstrates that the proposed DA mapping is general and can be used with various methods that consider the sample correlation matrix. We note that this example specifically considers our approach for the design of beam patterns (and not DoA estimation).}

\subsection{Induced Beamformer Spectra}
\label{Induces Beamformer Spectra}
This section considers a noise-only (null) scenario where the sample correlation matrix is $\mI$.
In this case, the matrix $\mE\mE^H$ can be viewed as the proposed DA applied to the sample correlation matrix $\mI$, which is white spatial noise. So, in a sense, it gives us the ``beamforming'' used by our DA. Additionally, it allows examining the steered response obtained by the adaptation independent of the source.

Another motivation for examining the spectrum of $\mE\mE^H$ is that it is the square norm of the vector used for computing the spectrum of the proposed approach with the DS spectrum, as detailed below.
The DS spectrum using the proposed approach in the null scenario is given by 
\begin{equation}
\label{eq: P_DS as norm of Ed}
    P_{\text{DS}}(\mE\mI\mE^H,\theta) = 
    \vd^H(\theta)\mE\mE^H\vd(\theta) =
    \|\mE^H\vd(\theta)\|^2.
\end{equation}
By denoting $\vu(\theta)=\mE^H\vd(\theta)$ the spectrum of the DS beamformer with the proposed approach for a sample correlation matrix $\sSigma$ is given by 
\begin{equation}
\label{eq: P_DS with u}
    P_{\text{DS}}(\mE\sSigma\mE^H,\theta) = 
    \vu^H(\theta)\sSigma\vu(\theta).
\end{equation}
So, (\ref{eq: P_DS as norm of Ed}) can be viewed as the norm of the vector that interacts with the correlation matrix to produce the spectrum of the DS beamformer. The matrix $\mE^H$ can be viewed as transporting the steering vector from the reference domain (consisting of steering vectors) to the operational domain. So, $\vu$ and $\sSigma$ in (\ref{eq: P_DS with u}) are in the same domain.

\begin{figure}
\centering
\subfloat[]{\includegraphics[width=0.5\linewidth]{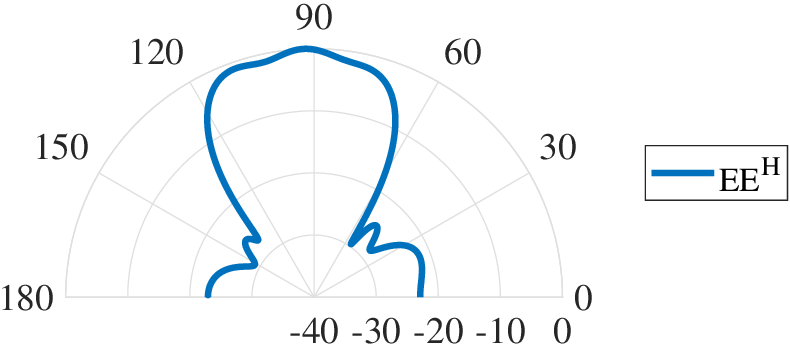}}
\subfloat[]{\includegraphics[width=0.5\linewidth]{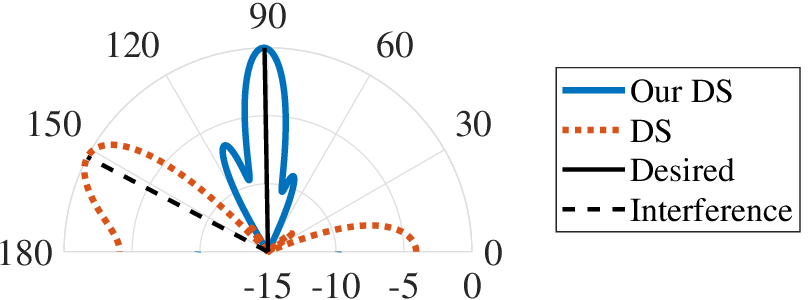}}
{\caption{Spectra for the setting shown in Figure \ref{fig: RF Room interference section} considering an interfering source with SIR $-20\text{dB}$. (a) The DS spectrum of $\mE\mE^H$. 
(b) Spectrum of the proposed approach with the DS beamformer (blue) and the typically used DS beamformer (red).}}
\label{fig: DS spectrum Two interf sections RF}
\end{figure}

In the first experiment, we consider the scenario shown in Figure \ref{fig: RF Room interference section} with two separate interference regions and SIR $-20\text{dB}$. Figure \ref{fig: DS spectrum Two interf sections RF}(a) presents the spectrum of the DS beamformer induced by $\mE\mE^H$ as the correlation matrix. We see one significant main lobe towards the region of interest. Directions from the two interference regions are significantly attenuated (over $20\text{dB}$). So, the proposed approach implicitly acts as a null sector towards the interference regions. Figure \ref{fig: DS spectrum Two interf sections RF}(b) presents the DS beamformer spectrum for a desired source contaminated by an interfering source. The spectrum of the proposed approach and the typically-used beamformers appear in blue and red, respectively. The directions of the desired and interfering sources appear in solid and dashed black, respectively. We see that the spectrum of the proposed approach has one main lobe, pointing at the desired source. Conversely, the spectrum of the typically used DS beamformer points toward the interfering source, with low power in the direction of the desired source. 

\begin{figure}[t]
  \begin{center}
\includegraphics[width=0.6\linewidth]{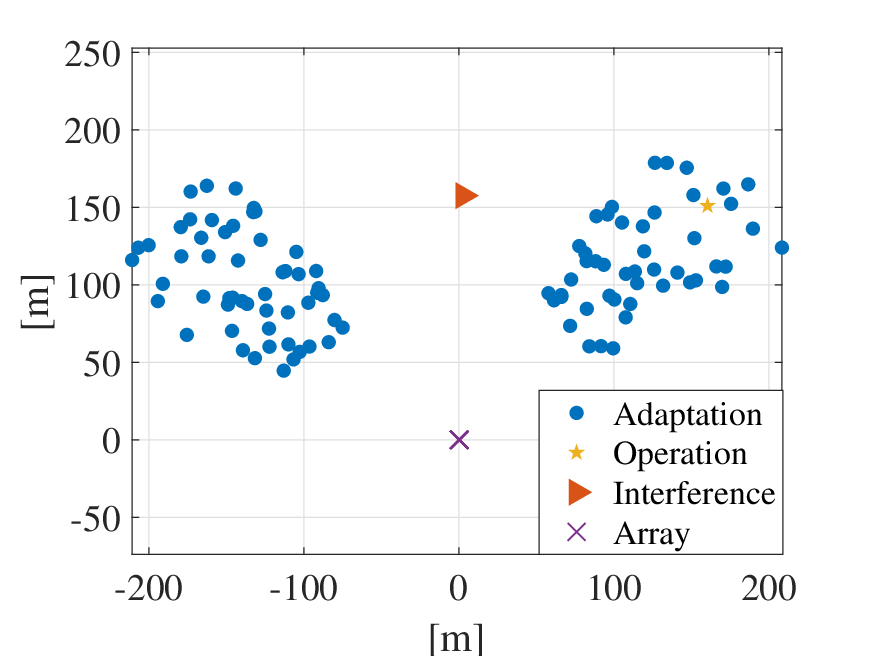}
\caption{The RF setting with two desired source sections, and one interfering source. }
\label{fig: RF Room two ROI}
\end{center}
\end{figure}

\begin{figure}[t]
\centering
\subfloat[]{\includegraphics[width=0.5\linewidth]{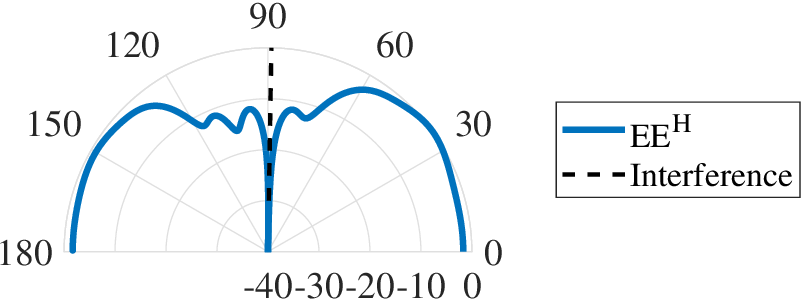}}
\subfloat[]{\includegraphics[width=0.5\linewidth]{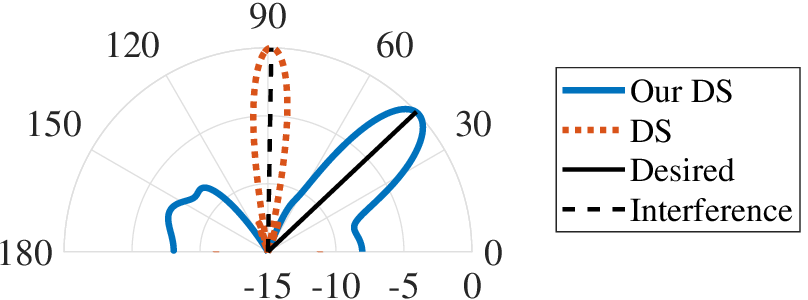}}
\caption{Spectra for the scenario shown in Figure \ref{fig: RF Room two ROI} with SIR $-20\text{dB}$. (a) The DS spectrum of $\mE\mE^H$ in the presence of an interfering source.
(b) The spectrum of the proposed approach with the DS beamformer (blue) and the typically used DS beamformer (red) in the presence of an interfering source.}
\label{fig: DS spectrum Two ROI RF}
\end{figure}

In the second experiment, we consider an RF setting with a region of interest comprising two sections, $\theta \in [30\degree,60\degree] \cup [130\degree,160\degree]$ and a single interfering source with SIR $-20\text{dB}$. Figure \ref{fig: RF Room two ROI} shows the scenario. It is the same as Figure \ref{fig: RF Room interference section} only for the current setting.
Figure \ref{fig: DS spectrum Two ROI RF}(a) presents the DS spectrum of $\mE\mE^H$. We see null at the direction of the interfering source. Furthermore, the spectrum attains high values in the direction of the region of interest, whereas the values are much lower in other directions. The proposed approach implicitly performs null steering in the direction of the interfering source. We emphasize that the interfering source's direction is unknown and need not be estimated at any stage.

\section{Conclusions}
\label{sec: Conclusion}
This work considers the DoA estimation of a desired source in the presence of interfering sources and multipath. We leverage prior knowledge of the known region of interest and the location of the phased array to compute a map applied to the received signal before the DoA estimation. The map is viewed as domain adaption, adapting the signal from a noisy domain to a reference environment, thereby mitigating the interfering sources and multipath. 
The proposed approach can be applied to various DoA estimation methods.
We show theoretically and empirically that the proposed approach improves DoA estimation using three commonly used beamformers for acoustic and RF signals.  

\bibliographystyle{IEEEtran}
\bibliography{DA_for_DoA}

\onecolumn
     
\section*{\LARGE Supplementary Material: Domain Adaptation for DoA Estimation in Multipath Channels with Interferences}

\normalsize

%

%

\vspace*{30px}


\setcounter{equation}{38}
\setcounter{table}{1}

%

%

%



\maketitle

\section{Proofs of Theoretical Results}
\label{Appendix: Proofs for the Theoretic Results}
\subsection{Proof of Proposition 1}
\begin{proof}
    \begin{equation}
    \begin{split}
        \mathrm{E}[\vy(l)\vy^H(l) ] &= 
        \mathrm{E}[\mathbf{E}\vz(l)\vz^H(l) \mathbf{E}^H ] \\ 
        &= 
        \mathbf{E}\mathrm{E}[\vz(l)\vz^H(l) ]\mathbf{E}^H  \\
        &= 
         \mathbf{\Sigma}_{\text{S}}
    ^{\frac{1}{2}}           \mathbf{\Sigma}_{\text{A}}   ^{-\frac{1}{2}}
     \mathbf{\Sigma}_{\text{A}}
    \left(     \mathbf{\Sigma}_{\text{S}}
    ^{\frac{1}{2}}         \mathbf{\Sigma}_{\text{A}}   ^{-\frac{1}{2}}\right)^H  \\
    &=
     \mathbf{\Sigma}_{\text{S}}
    \end{split}
\end{equation}
\end{proof}

\subsection{Proof of Proposition 2}
    \begin{proof} 
    We denote by $N_I(j)$ the number of interfering sources active during the transmission of the desired source associated with the $j$th position.
    So, $\tilde{N}_I = \sum_{j=1}^{N_A} N_I(j)$.
    
    Since the sources and the noise are uncorrelated, the following holds for the population correlation matrix.    
    \begin{equation}
    \begin{split}
      \mathbf{\Sigma}_j^{\text{A}} &=
      \mathbb{E}\left[
        \vz_j(l)\vz_j^H (l) \right]\\
        &=         
        \sigma_{s_j}^2 \vh_j\vh_j^H + 
        \sum_{i=1}^{N_I(j)}\sigma_{q_i}^2 \vg_i\vg_i^H + 
        \sigma_v^2\mI.       
    \end{split}
\end{equation}
Computing the mean results in
 \begin{equation}
    \begin{split}     
          \mathbf{\Sigma}_{\text{A}} &=              
    \frac{1}{N_{\text{A}}}\sum_{j=1}^{N_{\text{A}}}  \mathbf{\Sigma}^{\text{A}}_j
     \\
     &=      
    \frac{1}{N_{\text{A}}}\sum_{j=1}^{N_{\text{A}}} 
    \left(
        \sigma_{s_j}^2 \vh_j\vh_j^H + 
        \sum_{i=1}^{N_I(j)}\sigma_{q_i}^2 \vg_i\vg_i^H + 
        \sigma_v^2\mI
    \right) \\
    &=      
    \frac{1}{N_{\text{A}}}     
    \sum_{j=1}^{N_{\text{A}}}     
        \sigma_{s_j}^2
         \vh_j\vh_j^H      + 
        \sum_{i=1}^{\tilde{N}_I}\sigma_{q_i}^2 \vg_i\vg_i^H + 
        \sigma_v^2\mI.    
    \end{split}
\end{equation}

    \end{proof}

\subsection{Proof of Proposition 3}
\begin{proof}
    We examine the following matrix product.
\begin{equation}
\label{eq: E Gamma E expression in proof proposition 3}
\begin{split}
    \mE\mathbf{\Sigma}\mE^H &=
     \mathbf{\Sigma}_{\text{S}}^{\frac{1}{2}}
     \mathbf{\Sigma}_{\text{A}}^{-\frac{1}{2}}
    \mathbf{\Sigma}   
     \mathbf{\Sigma}_{\text{A}}^{-\frac{1}{2}}
      \mathbf{\Sigma}_{\text{S}}^{\frac{1}{2}} \\
    &=
     \mathbf{\Sigma}_{\text{S}}^{\frac{1}{2}}
     \mathbf{\Sigma}_{\text{A}}^{-\frac{1}{2}}
    \left( \mathbf{\Sigma}_{\text{A}} + \vd_S\vd_S^H\right)
     \mathbf{\Sigma}_{\text{A}}^{-\frac{1}{2}}
     \mathbf{\Sigma}_{\text{S}}^{\frac{1}{2}} \\
    &=
     \mathbf{\Sigma}_{\text{S}} +
    \mE
     \vd_S\vd_S^H
     \mE^H
\end{split}    
\end{equation}
Since the steering vectors $\vd_S$ and $\vd_I$ are i.i.d the following holds for the deterministic and known matrix $ \mathbf{\Sigma}_{\text{S}}$:
 \begin{equation}
     \mathbb{E}[\vd_S^H  \mathbf{\Sigma}_{\text{S}} \vd_S] =
     \mathbb{E}[\vd_I^H  \mathbf{\Sigma}_{\text{S}} \vd_I].
 \end{equation}
Consequently, to prove (\ref{eq: proposition 3 P_DS is better}) we focus on the second term of (\ref{eq: E Gamma E expression in proof proposition 3}) and show that
\begin{equation}
\begin{split}
     \mathbb{E}[\vd_I^H (
    \mE
     \vd_S\vd_S^H
     \mE^H )
     \vd_I]
     \le
     \mathbb{E}[\vd_S^H (
    \mE
     \vd_S\vd_S^H
     \mE^H )
     \vd_S]. 
\end{split}    
\end{equation}
We begin with the l.h.s.
\begin{equation}
\begin{split}             
      \mathbb{E}[      
    \vd_I^H     
    \mE
     \vd_S\vd_S^H       
    \mE^H
     \vd_I
     ] 
     &=     
      \mathbb{E}[
      \text{tr} (
    \vd_I^H     
    \mE
     \vd_S\vd_S^H       
    \mE
     \vd_I
     )]  \\     
     &=     
      \mathbb{E}[
      \text{tr} (    
     \vd_S\vd_S^H       
    \mE
     \vd_I\vd_I^H     
     \mE
     )] \\
     &=
     \text{tr} (
      \mathbb{E}[
    \vd_S\vd_S^H       
    \mE
     \vd_I\vd_I^H     
     \mE
     ]) \\
     &=
     \text{tr} (
      \mathbb{E}[
    \vd_S\vd_S^H]
    \mathbb{E}[\mE
     \vd_I\vd_I^H     
     \mE
     ])  \\
     &=
     \text{tr} (
      \mathbb{E}[
    \vd_I\vd_I^H]
    \mathbb{E}[\mE
     \vd_I\vd_I^H     
     \mE
     ]).
\end{split}    
\end{equation}
The equalities hold since $\vd_I$ and $\vd_S$ are i.i.d. We have that
\begin{equation}
\begin{split}             
      \mathbb{E}[      
    \vd_I^H     
    \mE
     \vd_S\vd_S^H       
    \mE^H
     \vd_I
     ] 
     &=          
     \text{tr} (
      \mathbb{E}[
    \vd_I\vd_I^H]
    \mathbb{E}[\mE
     \vd_I\vd_I^H     
     \mE
     ]) \\
     &\le 
     \text{tr} (
      \mathbb{E}[
    \vd_I\vd_I^H
    \mE
     \vd_I\vd_I^H     
     \mE
     ]) \\
     &=
     \mathbb{E}[      
    \vd_I^H     
    \mE
     \vd_I\vd_I^H       
    \mE
     \vd_I
     ] \\
     &=
     \mathbb{E}[      
    |\vd_I^H     
    \mE
     \vd_I|^2] \\
     &< 
     \mathbb{E}[      
    |\vd_S^H     
    \mE
     \vd_S|^2] \\
     &=
     \mathbb{E}[      
    \vd_S^H     
    \mE
     \vd_S\vd_S^H       
    \mE^H
     \vd_S
     ] .
\end{split}    
\end{equation}
The first inequality holds since 
\begin{equation}
    \mathbb{E}[
    \vd_I\vd_I^H]
    \mathbb{E}[ 
    \mE
     \vd_I\vd_I^H     
     \mE
     ]
     \preceq     
     \mathbb{E}[
    \vd_I\vd_I^H
    \mE
     \vd_I\vd_I^H     
     \mE
     ],
\end{equation}
which follows from the definition of the covariance of two matrices.
The second inequality is due to (\ref{eq: condition d_sEd_s > d_IEd_I}).
\end{proof}

\section{Additional Simulation Results}
 \label{Appendix: Additional Simulation Results}
We repeat the experiment of an active interfering source as described in the context of Figure \ref{fig: DoA performance for RF single interfering source}  in the paper 
for SNR $20\text{dB}$ and for varying SIR values. Table \ref{Table: RF SNR 20} presents the results. It is the same as Table \ref{Table: Acoustic one interference} 
but for the RF setting.
The proposed method leads to better DoA estimation accuracy and smaller IQR values for all beamformers and SIR values. Additionally, the higher the interfering source (lower SIR), the greater the improvement.
Next, we set the SIR to $-20\text{dB}$ and repeat the experiment for varying SNR values. Table \ref{Table: RF SIR -20} presents the results. The proposed approach leads to better accuracy and smaller IQR values for all the tested SNR values. The SNR has little effect on the performance of the typically used beamformers since the interference source is the main factor limiting the performance. In contrast, for the proposed approach, the higher the SNR is, the better the DoA estimations are.

\begin{table*}[h]
\caption{The median and IQR (in parenthesis) of the DoA estimation error for the RF setting for SNR $20$dB.}
\centering
\begin{tabular}{{c||c|c||c|c||c|c|}}
\multicolumn{1}{c||}{\bf Beamformer} 
&\multicolumn{1}{c|}{\bf DS}
&\multicolumn{1}{c||}{\bf Our DS }
&\multicolumn{1}{c|}{\bf MVDR}
&\multicolumn{1}{c||}{\bf Our MVDR}
&\multicolumn{1}{c|}{\bf MUSIC}
&\multicolumn{1}{c|}{\bf Our MUSIC}
 
 \\

 \hline 
  SIR=$0\text{dB}$   & 3.7 (37.4)     & \textbf{0.6} (0.8)     & 3.7 (36.9)     & \textbf{0.6} (0.7)     & \textbf{0.1} (26.7)     & {0.6} (0.8)  \\

  SIR=$-10\text{dB}$   & 34.3 (39.5)     & \textbf{0.6} (0.8)     & 34.3 (39.4)     & \textbf{0.6} (0.7)     & 28.5 (43.3)     & \textbf{0.6} (0.9)  \\

  SIR=$-20\text{dB}$   & 34.3 (39.4)     & \textbf{0.7} (1.0)     & 34.2 (39.4)     & \textbf{0.7} (1.0)     & 34.2 (39.5)     & \textbf{0.7} (1.1)  \\
\end{tabular}
\label{Table: RF SNR 20}
\end{table*}

\begin{table*}[h]
\caption{The median and IQR (in parenthesis) of the DoA estimation error for the RF setting for SIR $0\text{dB}$.}
\centering
\begin{tabular}{{c||c|c||c|c||c|c|}}
\multicolumn{1}{c||}{\bf Beamformer} 
&\multicolumn{1}{c|}{\bf DS}
&\multicolumn{1}{c||}{\bf Our DS }
&\multicolumn{1}{c|}{\bf MVDR}
&\multicolumn{1}{c||}{\bf Our MVDR}
&\multicolumn{1}{c|}{\bf MUSIC}
&\multicolumn{1}{c|}{\bf Our MUSIC}
 
 \\

  \hline 
  SNR=$20\text{dB}$   & 3.7 (37.4)     & \textbf{0.6} (0.8)     & 3.7 (36.9)     & \textbf{0.6} (0.7)     & \textbf{0.1} (26.7)     & {0.6} (0.8)  \\

  SNR=$10\text{dB}$   & 3.8 (37.4)     & \textbf{0.6} (0.8)     & 3.3 (37.5)     & \textbf{0.5} (0.7)     & 2.5 (35.5)     & \textbf{0.6} (0.8)  \\

  SNR=$0\text{dB}$   & 3.7 (37.4)     & \textbf{0.6} (0.9)     & 3.4 (37.5)     & \textbf{0.7} (1.0)     & 3.8 (37.1)     & \textbf{0.6} (0.9)  \\

\end{tabular}
\label{Table: RF SIR -20}
\end{table*}

%


\end{document}